\documentclass[lettersize,journal]{IEEEtran}
\usepackage{amsmath,amsfonts}
\usepackage{algorithmic}
\usepackage{algorithm}
\usepackage{array}
\usepackage{subcaption}
\usepackage{enumitem}
\usepackage{textcomp}
\usepackage{stfloats}
\usepackage{url}
\usepackage{verbatim}
\usepackage{graphicx}
\usepackage{cite}
\usepackage[hidelinks]{hyperref}
\hypersetup{
    colorlinks=true,
    citecolor=blue,   
    linkcolor=blue,   
    urlcolor=blue
}
\usepackage{footmisc}

\usepackage{booktabs}
\usepackage{multirow}
\usepackage{graphicx}
\usepackage{amssymb}
\usepackage{pifont}
\newcommand{\cmark}{\ding{51}}
\newcommand{\xmark}{\ding{55}}
\usepackage{threeparttable}
\usepackage{hyperref}

\newcommand{\supplementary}[1]{\textcolor{blue}{#1}}
\usepackage[table,dvipsnames]{xcolor}
\usepackage{xcolor}
\usepackage{booktabs}

\begin{document}

\title{SIMPLE: Semantic Structure-Aligned Prompt Learning for Open-Vocabulary Audio-Visual Event Localization}

\author{
Hanyu Xuan, 
Haoran Xu,
Min Yin,
Shuo Wang,~\IEEEmembership{Member,~IEEE,}
Cheng Zhao,
Kun Li,~\IEEEmembership{Member,~IEEE,}
Fei Wang,\\
Zhiliang Wu,~\IEEEmembership{Member,~IEEE,}
Guanghui Yue,~\IEEEmembership{Senior Member,~IEEE,}
Hehe Fan,~\IEEEmembership{Senior Member,~IEEE}

\thanks{This work was supported in part by the Natural Science Foundation of Anhui Province (No.2308085QF221), 
in part by the National Natural Science Foundation of China (No.62302006 and No.12271002). (\emph{Corresponding author: Zhiliang Wu})}
\thanks{Hanyu Xuan, Haoran Xu and Min Yin are with the School of Big Data and Statistics,
Anhui University, Hefei 230039, China.}
\thanks{Shuo Wang is with the School of Information Science and Technology, University of Science and Technology of China, Hefei, China.}
\thanks{Cheng Zhao and Guanghui Yue are with the Shenzhen University Medical School, Shenzhen University, Shenzhen 518060, China (e-mail: chengzhao@szu.edu.cn and yueguanghui@szu.edu.cn).}
\thanks{Kun Li is with the College of Information Technology, United Arab Emirates University, Abu Dhabi, 15551, United Arab Emirates (e-mail: kun.li@uaeu.ac.ae).}
\thanks{Fei Wang is with the Institute of Artificial Intelligence, Hefei Comprehensive National Science Center, Hefei, 230026, China (e-mail: jiafei127@gmail.com). }
\thanks{Zhiliang Wu is with the College of Computing and Data Science, Nanyang Technological University, Singapore 639798, Singapore (e-mail: zhiliang.wu@ntu.edu.sg).}
\thanks{Hehe Fan is with the School of Artificial Intelligence, Zhejiang University, Hangzhou 310007, China (e-mail: hehefan@zju.edu.cn).}
}

\markboth{IEEE Transactions on Image Processing,~Vol.~XX, No.~XX, November~2026}%
{Shell \MakeLowercase{\textit{et al.}}: A Sample Article Using IEEEtran.cls for IEEE Journals}

\IEEEpubid{}
\maketitle

\begin{abstract}
Open-vocabulary audio-visual event localization aims to infer explicit segment-level categories for both seen and unseen audio-visual events.
Existing solutions rely heavily  on hand‑crafted prompt templates, whose predictions are highly sensitive to subtle wording variations.
Learnable prompts remove this manual tuning burden,
but when trained solely on seen events, 
they tend to overfit to seen categories
and distort the inter-event semantic topology required for generalization to unseen categories, 
a failure we term semantic topology drift.
In this paper, we propose SIMPLE, a \underline{S}emantic Structure-al\underline{I}gned Pro\underline{MP}t \underline{LE}arning framework that aligns cross-category semantic topology between learnable and hand-crafted prompts across asynchronous audio and visual streams.
Specifically, 
our SIMPLE combines  Learnable Continuous Prompt Tuning (LCPT) with 
Semantic Topology Consistency Regularization (STCR).
The LCPT automates prompt engineering through data-driven  optimization,
while the STCR module aligns the category-wise similarity distributions of learnable and hand-crafted prompts 
under random category subset sampling to alleviate semantic topology drift.
Comprehensive experiments and analyses on the OV-AVEBench demonstrate that our SIMPLE framework achieves state-of-the-art performance on the seen and unseen splits, while using roughly half the trainable parameters and roughly one-third the training time.
\end{abstract}

\begin{IEEEkeywords}
open-vocabulary,
prompt learning,
cross-modal alignment,
semantic topology drift.
\end{IEEEkeywords}

\section{Introduction}

\IEEEPARstart{H}{umans} can instinctively associate  visual appearances with their corresponding sounds and distinguish between them through multi-sensory perception
\cite{treichler1967training,wen2009multisensory}.
This remarkable cross-modal capability enables humans to perceive not only which event is occurring, but also precisely when the event occurs.
Inspired by this human perceptual ability,
the task of Audio-Visual Event Localization (AVEL)
\cite{tian2018audio,Wu_2019_ICCV,xu2020cross,Duan_2021_WACV,xuan2021discriminative,xuan2024robust,shen2023fine,Zhao_2024,shaar2025adapting,xuan2022proposal,Zhang2025AVEL,Xia_2022_CVPR,zhou2022contrastive,liu2024masked,wang2024context,liu2025scess,Sun_2026_CVPR,wu2022span,yu2022mm,xuan2022active,zhang2025multi,ZhangYMR25,zhou2025dense,xuan2026task}
has been proposed to simultaneously identify the semantic categories and temporal boundaries of audio-visual events within untrimmed videos.
In contrast to uni-modal machine perception
\cite{tan2026selective},
AVEL requires joint modeling of asynchronous audio and visual streams to predict both event categories and their temporal intervals,
making it a fundamental yet challenging problem in multi-modal video understanding
\cite{xing2026ma,zhang2023pfta,li2026ma,wu2026dlvinet}.

\begin{figure}[t]
\centering
\includegraphics[width=1.0\linewidth]{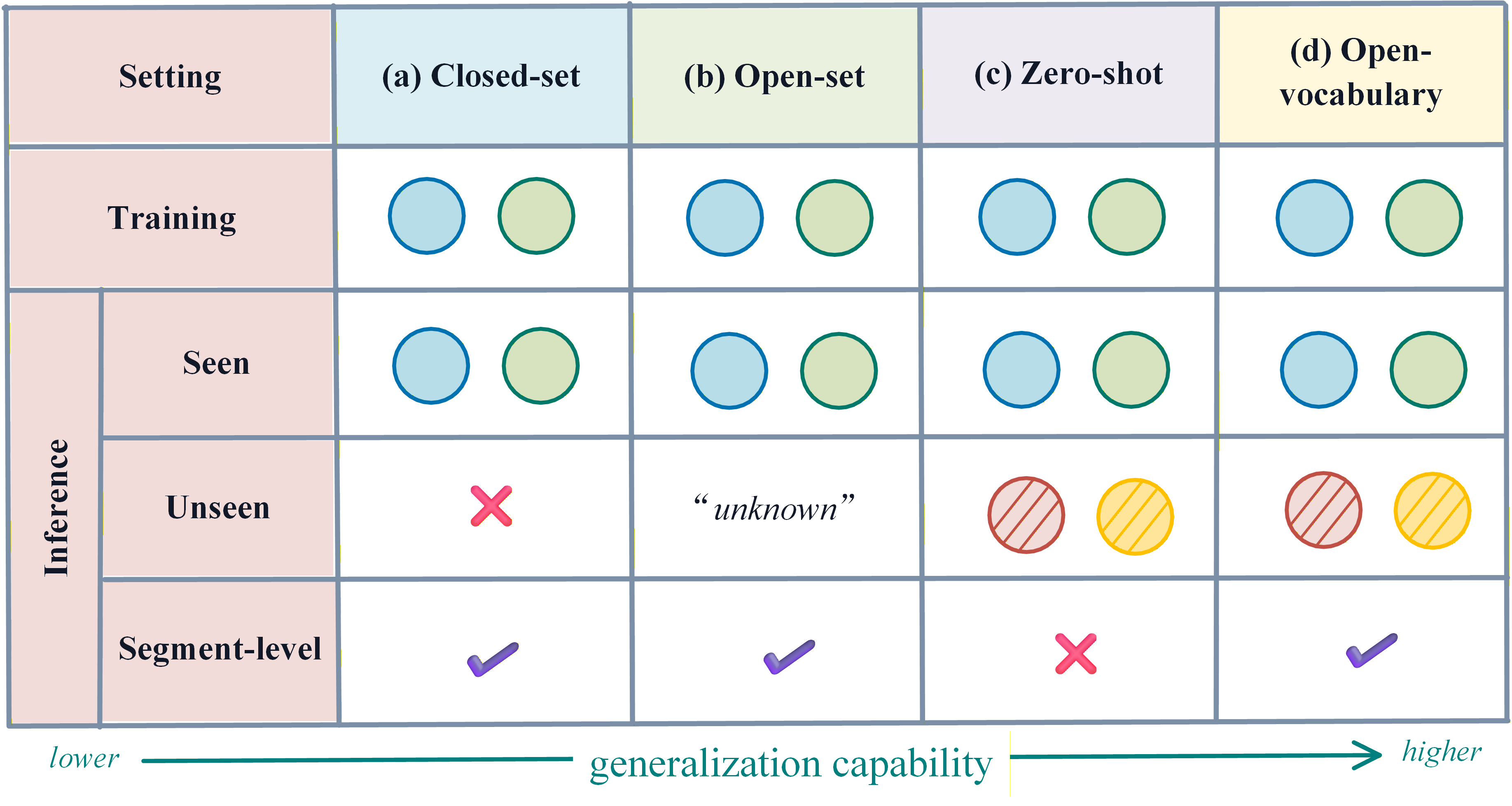}
\caption{Conceptual comparison of closed-set, open-set, zero-shot, and open-vocabulary settings for Audio-Visual Event Localization (AVEL) task.}
\label{fig1}
\end{figure}

Since the pioneering work of Tian et al.~\cite{tian2018audio},
a wide range of methods \cite{Wu_2019_ICCV,xu2020cross,Duan_2021_WACV,xuan2021discriminative,shen2023fine,Zhao_2024,shaar2025adapting,Zhang2025AVEL,Xia_2022_CVPR,zhou2022contrastive,liu2024masked,wang2024context,liu2025scess,Sun_2026_CVPR,wu2022span,yu2022mm,zhang2025multi,ZhangYMR25,zhou2025dense} have been proposed.
Although these methods have achieved promising performance,
they are inherently limited to closed-set task settings.
As illustrated in Fig.~\ref{fig1}(a),
such methods can recognize only the audio-visual event categories
encountered during training, which severely hinders their generalization to
test videos containing unseen audio-visual event categories.

To mitigate this limitation, 
Yu et al. \cite{yu2024openave} investigated the AVEL task under an open-set setting. 
To the best of our knowledge, 
this remains the only study dedicated to this particular problem setting.
Specifically,
open-set AVEL is capable of inferring novel audio-visual events not observed  during training
and recognizing these unseen events as the “\textit{unknown}” category,
instead of assigning them to pre-defined known event categories.
Nevertheless,
as depicted in Fig.\ref{fig1}(b),
this open-set approach \cite{yu2024openave} is unable to infer the fine-grained semantic categories of these unseen events.

To further alleviate this bottleneck,
several studies~\cite{Mazumder_2021_WACV,mercea2022ECCV,Hong_2023_ICCV,mo24avzsl,MA2026133693} have explored the AVEL task
under the zero-shot setting.
This setting aims to predict explicit categories for both seen and
unseen events during inference, rather than assigning unseen events
a generic “\textit{unknown}” label, as in open-set AVEL.
Notably, as shown in Fig.~\ref{fig1}(c),
existing zero-shot AVEL methods~\cite{Mazumder_2021_WACV,mercea2022ECCV,Hong_2023_ICCV,mo24avzsl,MA2026133693} operate at the whole-video level, predicting only a single event category per video.
Consequently, they forego fine-grained temporal boundary estimation and thus cannot localize audio-visual events to their respective temporal segments.

To bridge this research gap,
Zhou et al.~\cite{zhou2025towards} extended the AVEL task to the open-vocabulary scenario.
As represented in Fig.~\ref{fig1}(d),
open-vocabulary AVEL aims to temporally localize audio-visual events
and infer explicit semantic categories for both seen and unseen events.
Compared with the closed-set setting,
open-vocabulary AVEL exhibits stronger generalization to unseen categories.
Distinct from the open-set setting \cite{yu2024openave},
it predicts concrete semantic labels for unseen events
instead of assigning them a generic \textit{unknown} tag.
In contrast to the whole-video zero-shot setting~\cite{Mazumder_2021_WACV,mercea2022ECCV,Hong_2023_ICCV,mo24avzsl,MA2026133693},
open-vocabulary AVEL supports fine-grained, segment-level temporal localization of audio-visual events.

\begin{figure}[t]
\centering
\includegraphics[width=0.7\linewidth]{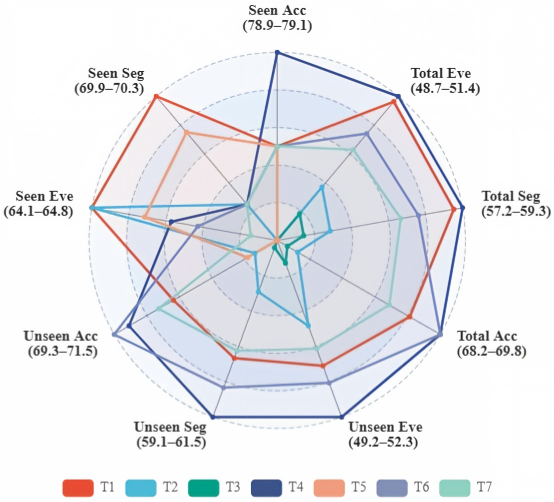}
\caption{Performance comparison across various hand-crafted templates on the open-vocabulary AVEL benchmark~\cite{zhou2025towards}.}
    \begin{tablenotes}[flushleft]
    \scriptsize
    \item
    $T_1$ \cite{kurzendorfer2024audio}: a video of $[\text{event}_i]$;
     $T_2$ \cite{hanif2024palm}: an audio recording of $[\text{event}_i]$;
    $T_3$ \cite{hanif2024palm}: the sound of $[\text{event}_i]$;
     $T_4$ \cite{hanif2024palm}: a video clip showing $[\text{event}_i]$;
    $T_5$ \cite{zhou2025towards}: an audio-visual event of $[\text{event}_i]$;
     $T_6$ \cite{jung2025learning}: a scene with $[\text{event}_i]$ sound and visual;
    $T_7$ \cite{jung2025learning}: $[\text{event}_i]$ sound with visual context.
  \end{tablenotes}
\label{fig:templates}
\end{figure}

Specifically, to enable open-vocabulary recognition,
Zhou et al.~\cite{zhou2025towards} constructed hand-crafted textual
descriptions for each candidate audio-visual event category.
These fixed textual prompts are then fed into the text encoder
of a pre-trained Multi-modal Large Language Model (MLLM)~\cite{xu2026visulogic,zheng20263d,wu2026scentgen,zhou2025dropping,gao2025feature,gao2025dt}.
Final category predictions are generated via dynamic similarity matching across audio, visual, and text embeddings,
enabling the model to classify arbitrary event categories
\cite{duan2026structure}. 
Nevertheless,
we argue that this strategy rooted in hand‑crafted prompt templates for in‑context learning \cite{de2026context}
exhibits several inherent limitations.
As demonstrated in Fig.\ref{fig:templates},
the predictions generated using such prompts are highly sensitive
to minor wording variations, which can lead to substantial
performance fluctuations~\cite{fillioux2026soc}.
Despite extensive trial-and-error tuning,
identifying a universally optimal prompt template remains an intractable challenge.
Moreover, these static textual descriptions lack the flexibility to adaptively capture complex audio-visual semantics
\cite{zhou2022learning}.

To address these limitations,
we introduce Learnable  Continuous Prompt Tuning (LCPT).
Building upon CoOp~\cite{zhou2022learning},
LCPT generalizes continuous prompt tuning
from uni-modal image classification  
to bi-modal temporal event localization
in the open-vocabulary scenario,
while freezing the parameters of pre-trained MLLMs.
It parameterizes the context components of prompts as learnable continuous vectors, rather than discrete text tokens whose embeddings remain fixed during downstream prompt tuning. 
This design automates prompt engineering via  data-driven gradient-based optimization, and harnesses the rich pre-trained knowledge encoded within MLLMs for task-specific cross-modal alignment~\cite{liang2026prompt}.
Unlike CoOp~\cite{zhou2022learning},
which optimizes the context exclusively for visual input,
our LCPT learns a unified context across asynchronous audio and visual streams.

However, under the open-vocabulary setting,
learnable prompts trained exclusively on seen event categories
exhibit limited generalization ability to
unseen event categories.
Through a systematic analysis of failure cases,
we observe that this generalization deficit stems from overfitting of
the text embeddings produced by  learnable prompts to the semantic distribution of target seen categories during prompt tuning, 
which distorts the native inter-class semantic topology encoded within pre-trained MLLMs.
We formally term this phenomenon as semantic topology drift.
As illustrated in Fig.\ref{fig:drift},
such biased text embeddings disrupt the
consistent similarity structure between seen and unseen event categories at inference time.
Once cross-category semantic associations are incorrectly formed, they act as low-quality classifiers and degrade
open-vocabulary generalization.
Extended analysis are provided in  Sect.~\ref{Exper} and \supplementary{Supplementary Materials}.

In contrast, hand-crafted prompts maintain stable and consistent semantic distributions across different categories,
as they are free from task-specific parameter updates and can faithfully preserve the intrinsic semantic topology embedded in pre-trained MLLMs. 
Motivated by these crucial observations, 
we introduce Semantic Topology Consistency Regularization (STCR) to mitigate semantic topology drift and boost cross-category generalization capability.
Specifically, 
we first leverage Qwen2-7B~\cite{Yang2024Qwen2TR}
to generate category names for synthetic novel events,
which serve as simulated unseen events throughout training. 
Notably, all synthetic categories are generated entirely without access to test-set information
and strictly verified to be fully disjoint from the benchmark’s real unseen categories,
eliminating any risk of test-set information leakage.
For both seen and synthetic novel categories,
the STCR module aligns the pairwise similarity relationships between text embeddings from learnable prompts with those from hand-crafted prompts.
By constraining and preserving the intrinsic cross-category semantic topology, 
STCR facilitates effective semantic transfer from seen categories to simulated unseen categories, 
which naturally generalizes to real unseen categories during inference and thus boosts the generalization performance of open-vocabulary AVEL task.

We name our framework \textbf{S}emantic Structure-al\textbf{I}gned Pro\textbf{MP}t \textbf{LE}arning (SIMPLE), 
as it aligns the semantic topology structures between learnable and hand‑crafted text embeddings to preserve meaningful inter‑event relationships.
The SIMPLE framework consists of 
an LCPT module for automated prompt engineering
and an STCR component for mitigating the semantic topology drift.
The main contributions are summarized as follows:
\begin{itemize}
    \item To the best of our knowledge, this work constitutes the first attempt to apply learnable prompt tuning to open-vocabulary audio-visual event localization,
    substituting labor-intensive hand-crafted template  tuning with  data‑driven gradient-based optimization. 
    We further reveal a critical limitation:
    learnable prompts tend to distort the inter-event semantic topology in this task.
    
    \item We design a semantic structure-aligned prompt learning framework tailored to open-vocabulary AVEL task. It combines   learnable continuous prompt tuning with semantic topology consistency regularization.

    
    

    \item We achieve the state-of-the-art performance on both the seen and unseen splits of OV-AVEBench with substantially fewer trainable parameters and lower training cost.
\end{itemize}

\section{Related Work}

\subsection{Audio-Visual Event Localization}
AVEL aims to temporally localize and classify events in untrimmed videos by jointly exploiting audio and visual modalities. 
As shown in Fig.~\ref{fig1}, 
existing AVEL methods fall into four distinct settings: 

\subsubsection{Closed-set AVEL}
Since Tian et al. \cite{tian2018audio}
first formalized AVEL task,
most existing studies have focused on the closed-set assumption.
Existing AVEL approaches can be broadly classified into three categories:

\textit{Audio-visual fusion}:
A core challenge in AVEL lies in effectively integrating complementary information across audio-visual modalities. Wu et al.~\cite{Wu_2019_ICCV} proposed dual-attention matching to achieve bidirectional alignment of audio and visual features. Xu et al.~\cite{xu2020cross} integrated audio-guided spatial-channel attention~\cite{wu2024waveformer} with relation-aware aggregation to emphasize event-relevant visual regions. Duan et al.~\cite{Duan_2021_WACV} iteratively fed the fused audiovisual representations into both modalities to enable progressive multi-modal fusion. Xuan et al.~\cite{xuan2021discriminative} designed a discriminative cross-modality attention~\cite{yue2025text} network to address temporal inconsistency in AVEL. Shen et al.~\cite{shen2023fine} further explored fine-grained audiovisual interactions to facilitate adaptive feature fusion. Zhao et al.~\cite{Zhao_2024} incorporated event-level semantic information to strengthen multi-modal event representation and modeling. Shaar et al.~\cite{shaar2025adapting} developed a training-free score-level fusion framework equipped with a within-video label-shift adjustment. 
Zhang et al.~\cite{Zhang2025AVEL} modeled bidirectional temporal dependencies and transferred CLIP-based image-text knowledge to the audiovisual event domain.

\begin{figure}[t]
\centering
\includegraphics[width=0.8\linewidth]{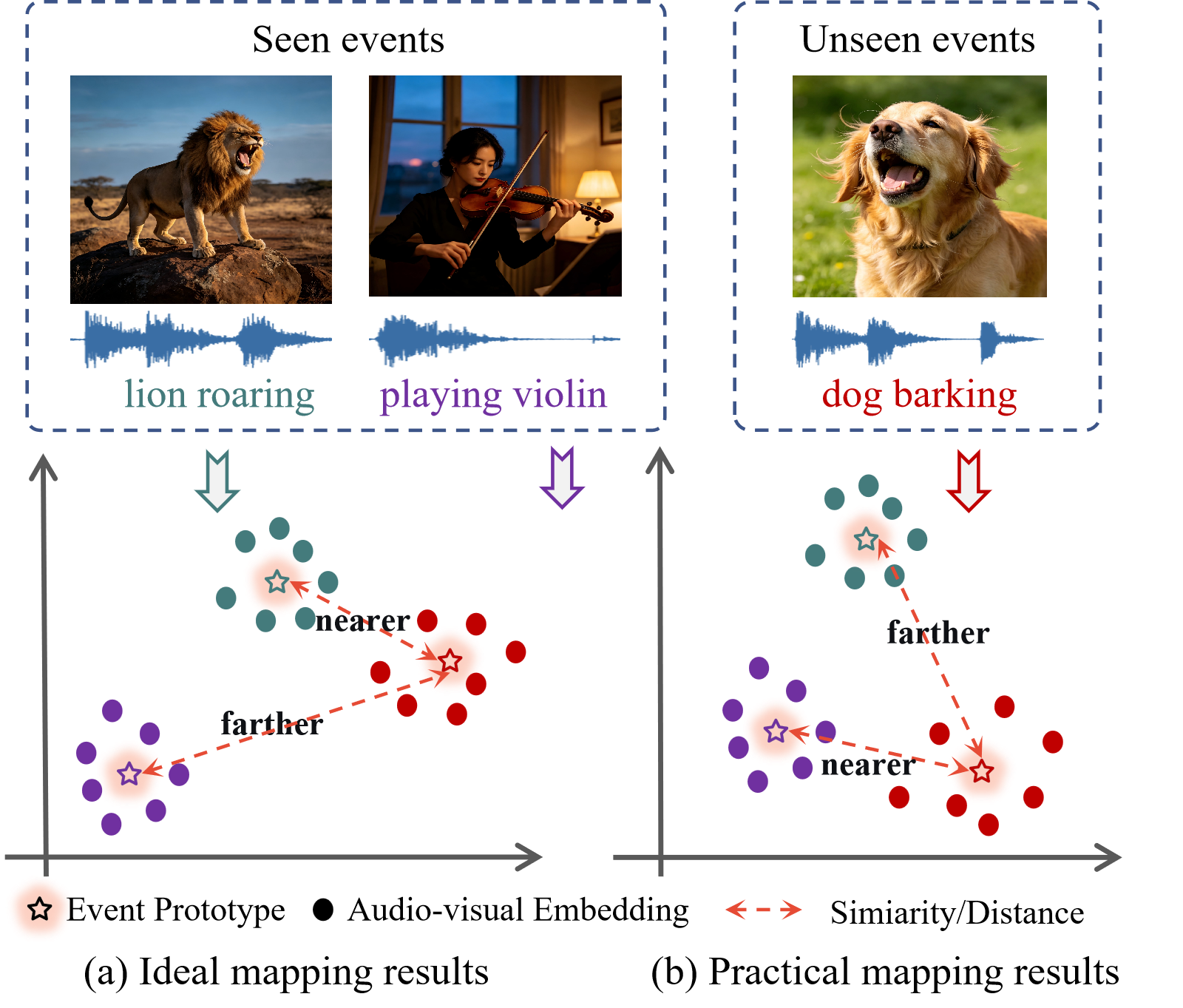}
\caption{Illustration of semantic topology drift induced by learnable prompts.
The cross‑modal mapping learned from seen‑event samples mis-aligns the embeddings of unseen events,
leading to distorted inter‑event semantic topology.
}
\label{fig:drift}
\end{figure}

\textit{Interference suppressing}:
A complementary research thrust aims to disentangle foreground events from background interference.
Xia et al. \cite{Xia_2022_CVPR} applied dual-level cross-modal suppression across temporal and event dimensions.
Zhou et al. \cite{zhou2022contrastive} identified highly relevant audio-visual pairs via similarity-based propagation to learn more discriminative event representations.
Departing from conventional strategies that treat backgrounds as pure noise, Liu et al.~\cite{liu2024masked} introduced a masked co-attention mechanism that explicitly masks out background-modality features during cross-modal interaction.
Wang et al.~\cite{wang2024context} built a context-aware proposal boundary network with structural consistency constraints to suppress noisy background proposals and refine temporal localization.
Liu et al.~\cite{liu2025scess} combined semantic consistency enhancement with segment selection to suppress background disturbances.
Sun et al. \cite{Sun_2026_CVPR} integrated explicit background semantics via multi-modal fusion of audio, visual, and text modalities.

\textit{Multi-modal temporal modeling}:
Another research stream focuses on precise temporal prediction and coherent segment modeling for audio‑visual events.
Wu et al. \cite{wu2022span} exploited global and local temporal cues through dual-attention matching. 
Yu et al. \cite{yu2022mm} further introduced a multi-modal temporal pyramid to capture event semantics across multiple temporal scales and accommodate events with varied durations. 
Zhang et al. \cite{zhang2025multi} modeled continuity relations among adjacent event segments to improve temporal coherence.
Zhang et al.\cite{ZhangYMR25} further enhanced event-level temporal consistency through multi-modal feature fusion and adaptive continuity sampling. 
Zhou et al. \cite{zhou2025dense} extended AVEL to dense localization within long untrimmed videos, 
leveraging cross-modal consistency and multi-temporal granularity collaboration to achieve finer‑grained timestamp prediction.

\subsubsection{Open-set AVEL}
Under the open-set setting, models are required not only to recognize events from known categories, 
but also to identify unseen events and treat them as “unknown” rather than forcibly assigning them to predefined classes. 
To date, this setting remains largely unexplored in the AVEL literature. 
To the best of our knowledge, 
Yu et al.~\cite{yu2024openave} presented the sole work addressing this challenge.
They proposed an evidential learning-based framework that quantifies classification uncertainty,
enabling effective separation of unknown events from both known events and background interference.

\subsubsection{Zero-shot AVEL}
The zero-shot setting aims to predict explicit categories for both seen and unseen events at inference time.
Mazumder et al.~\cite{Mazumder_2021_WACV} reconstructed semantic label features from multi-modal embeddings to facilitate knowledge transfer to unseen categories. 
Mercea et al.~\cite{mercea2022ECCV} further introduced temporal cross-attention~\cite{wu2025bvinet} to exploit audio-visual correspondence over time for zero-shot recognition. 
Hong et al.~\cite{Hong_2023_ICCV} embedded audio-visual semantics into hyperbolic space to better capture the hierarchical structure of unseen categories. 
Mo et al.~\cite{mo24avzsl} adopted supervised contrastive learning to directly align audio-visual features with textual semantics. 
Ma et al.~\cite{MA2026133693} addressed fusion degradation and modality imbalance through semantic-enhanced pseudo-feature generation and modality-adaptive fusion. 
Despite these advances, existing methods primarily target video-level recognition.

\subsubsection{Open-vocabulary AVEL}
Open‑vocabulary AVEL defines a more general paradigm, where models are required to temporally localize events and predict semantic labels for both seen and unseen event categories. 
Zhou et al.~\cite{zhou2025towards} pioneered this task and constructed the OV‑AVEBench benchmark. 
Yu et al.~\cite{Yu2025OVDAVELTO} further extended this paradigm to dense event localization within long untrimmed videos.  
They presented a one‑stage framework that jointly generated foreground proposals and conducted semantic‑aware classification for unseen events. 
Yang et al.~\cite{yang2026hierarchical} exploited heterogeneous‑graph‑based hyperbolic representations to model segment‑ and video‑level hierarchical semantics, improving cross‑modal consistency and generalization for unseen categories.
Cho et al.~\cite{jocompositional} explored decision‑layer adaptation with frozen multi-modal encoders, adopting agent‑discovered symbolic rules and video‑aware inference policies to boost seen‑unseen classification calibration.

\subsection{Prompt‑based Learning}
Prompt‑based learning \cite{lei2024prompt} adapts frozen MLLMs,
such as CLIP~\cite{radford2021learning},
via task-oriented prompt conditioning, 
covering both training‑free in‑context inference and learnable soft‑prompt tuning schemes.

\subsubsection{In‑context Learning}
Several studies~\cite{zhou2025towards,kurzendorfer2024audio,hanif2024palm,pratt2023does}
employ hand-crafted prompts for in-context learning~\cite{de2026context} to establish semantic connections across text, audio, and visual modalities. 
As summarized in the notes of Fig.\ref{fig:templates}, these hand-crafted templates cover single-modality descriptions for audio or video, as well as cross-modal formulations that describe audio-visual events and their contextual co-occurrence. 
Such fixed textual templates serve as guidance for MLLMs to perform semantic matching and localization of target events \([\text{event}_i]\).
Nevertheless,
all hand-crafted templates suffer from a shared limitation: 
their performance depends on how well their wording matches the encoder's pre-training distribution. 
Manually engineered prompts are inherently brittle, and no single template yields optimal performance across all event categories and datasets.

\subsubsection{Prompt Tuning}
Prompt tuning~\cite{li2025survey} optimizes lightweight soft-prompt embeddings while freezing pre-trained encoders, 
providing a parameter-efficient paradigm for adapting MLLMs to downstream tasks.
Zhou et al.~\cite{zhou2022learning} 
replaced hand-crafted prompts with a set of learnable context vectors optimized on few-shot labeled samples.
While achieving substantial performance gains on seen classes, this approach suffers from overfitting to training classes and degraded generalization on unseen classes.
Zhou et al.~\cite{zhou2022conditional} 
conditioned prompts on image features to generate image-adaptive prompts, which yield improved generalization toward unseen classes.
Khattak et al.~\cite{khattak2023self,khattak2023maple}
extended adaptability to downstream tasks by adding prompts to both text and image inputs.
Hanif et al.~\cite{hanif2024palm}
learned prompt context within the text feature space instead of the input token embedding space.

A variety of regularization techniques have been investigated to learn generalizable prompts.
Zhu et al.~\cite{zhu2023prompt} avoided gradient conflicts with CLIP's general knowledge via orthogonal projection of prompt-tuning gradients.
Yao et al.~\cite{yao2023visual} minimized the distance between text embeddings from learnable and hand-crafted prompts to preserve the general knowledge.
Zhang et al.~\cite{zhang2024generalizable} regarded hand-crafted and learnable prompts as dual views and fused task-specific and general semantics via a mutual-information estimator.
Chen et al.~\cite{Chen2024_b9603de9} maximized the Pearson correlation coefficient between pairwise similarity matrices of original and learnable text embeddings to preserve class topology.
Bulat et al.~\cite{bulat2023lasp} introduced a text-to-text cross-entropy loss to align embeddings from learnable and hand-crafted prompts of the same class.
Cho et al.~\cite{Cho_2023_ICCV} enforced uniform text embedding distribution on a hyper-sphere to reduce inter-class overlap and cluster same-class image embeddings for cross-modal alignment.
Jung and Lee \cite{jung2025learning} introduced similarity alignment regularization  to improve the generalization of CLIP-based prompt learning by preserving inter-class semantic relationships.

\section{Methodology}

\subsection{Preliminaries}

Consider an audible video $\{A_t, V_t\}_{t=1}^T$ consisting of $T$ consecutive and non-overlapping temporal segments, 
where $A_t$ and $V_t$ denote the audio and visual components of the $t$-th segment, respectively. 
Let $\textbf{\textit{Y}}=\{\textbf{\textit{y}}_t\}_{t=1}^T \in \mathbb{R}^{T \times C}$ denote the ground truth (GT), where $C$ is the total number of audio-visual event categories plus a background class. 
Open-vocabulary AVEL seeks to temporally localize  audio-visual events within a video and predict their corresponding categories $\hat{\textbf{\textit{y}}}_t$, such that:
\begin{equation}
   \hat{\textbf{\textit{y}}}_t \in \textbf{\textit{Y}}^s \cup \textbf{\textit{Y}}^u \subseteq \mathbb{R}^{C^s+C^u}, \quad \text{where } \textbf{\textit{Y}}^s \cap \textbf{\textit{Y}}^u = \emptyset.
\end{equation}
Here, $\textbf{\textit{Y}}^s \in \mathbb{R}^{C^s}$ indicates $C^s$ seen event categories available during training, 
while $\textbf{\textit{Y}}^u \in \mathbb{R}^{C^u}$ corresponds to the $C^u$ unseen event categories, 
which are inaccessible  during training time yet must be inferred at testing time.

To facilitate generalization toward unseen categories,
we leverage the zero-shot capabilities  of ImageBind \cite{girdhar2023imagebind}.
Specifically,
we adopt its audio encoder $\boldsymbol{\phi}_{a}(\cdot)$, visual encoder $\boldsymbol{\phi}_{v}(\cdot)$, and text encoder $\boldsymbol{\phi}_{txt}(\cdot)$ as our backbone. 
We extract the segment-wise audio and visual features, 
denoted as
$\textbf{\textit{F}}^a = \{ \textbf{\textit{a}}_t \}_{t=1}^T$ and
$\textbf{\textit{F}}^v = \{ \textbf{\textit{v}}_t \}_{t=1}^T$,
where $\textbf{\textit{a}}_t, \textbf{\textit{v}}_t \in \mathbb{R}^d$.
Subsequently,
a lightweight temporal Transformer layer
is applied to each modality independently to capture intra-modal sequential dependencies.

Zhou et al.~\cite{zhou2025towards}
construct tailored  textual descriptions $\text{``}[\text{event}_i] \text{''}$ 
as  hand-crafted prompt templates
for both seen and unseen event categories.
Similarly, 
subsequent studies  
\cite{kurzendorfer2024audio,hanif2024palm,jung2025learning,pratt2023does}
have proposed diverse hand-crafted prompt templates.
During the training phase,
these textual descriptions are fed into the text encoder $\boldsymbol{\phi}_{txt}(\cdot)$ to produce a sentence-level embedding:
\begin{equation}
  \textbf{\textit{h}}_i = \boldsymbol{\phi}_{txt}
  (\text{``}~ T_i ~\text{''}),
  \label{Eq2}
\end{equation}
where $\textbf{\textit{h}}_i \in \mathbb{R}^{d}$.
We then collect the embeddings of all $C^s$ seen event categories to form the set 
$\textbf{\textit{F}}^h = \{ \textbf{\textit{h}}_i \}_{i=1}^{C^s}$.

We conduct  performance comparisons across seven hand‑crafted prompts $T_1,\dots,7$
on the OV‑AVEBench benchmark~\cite{zhou2025towards} for open-vocabulary AVEL.
As detailed  in Fig.\ref{fig:templates},
model performance is highly sensitive to subtle wording variations,
making it non-trivial to identify an optimal template even after exhaustive trial‑and‑error tuning.
Further experimental observations and in‑depth analyses are provided in Sect.~\ref{Exper}.
Furthermore,
we argue that these static templates
lack the semantic flexibility  required  to  adapt to complex audio‑visual semantics under open-vocabulary scenarios \cite{zhou2022learning}.

\subsection{Semantic Structure-aligned Prompt Learning}

\subsubsection{Learnable Continuous Prompt Tuning}
\label{sect:LCPT}
To overcome the inherent  limitations of static hand‑crafted prompts,
we introduce an LCPT module to  automate prompt engineering
while freezing  the pre-trained parameters of
text encoder $\boldsymbol{\phi}_{txt}(\cdot)$.
Specifically,
LCPT introduces $P$ learnable context vectors
$\{\boldsymbol{p}_i\}_{i=1}^P$ and constructs the prompt sequence as
$[\boldsymbol{p}_1, \dots, \boldsymbol{p}_{_P}, \boldsymbol{c}_i]$,
where $\boldsymbol{c}_i$ indicates the token embedding of the $i$-th event category.
These continuous context vectors 
enable the model  to produce  task-adaptive 
text embeddings in a data-driven manner:
\begin{equation}
  \boldsymbol{l}_i = \boldsymbol{\phi}_{txt}
  ([\boldsymbol{p}_1, \dots, \boldsymbol{p}_{_P}, \boldsymbol{c}_i]).
  \label{Eq3}
\end{equation}

Compared with discrete hand-crafted prompts,
our LCPT possesses three core merits.
First, its unconstrained continuous embedding space
captures fine-grained, task-adaptive  audio-visual semantic  patterns that cannot be modeled by rigid discrete prompt templates.
Second, learnable vectors support end-to-end gradient-based optimization,
supporting flexible and efficient prompt adaptation.
Third, such optimizable prompt embeddings produce stable semantic prototypes,
which facilitate consistent and effective cross-modal matching.
These advantages collectively enhance the open-vocabulary generalization capability of our model, 
as further validated by detailed experimental results and analyses in Sect.~\ref{Exper}.

\begin{figure}[!t] 
    \centering \includegraphics[width=0.99\linewidth]{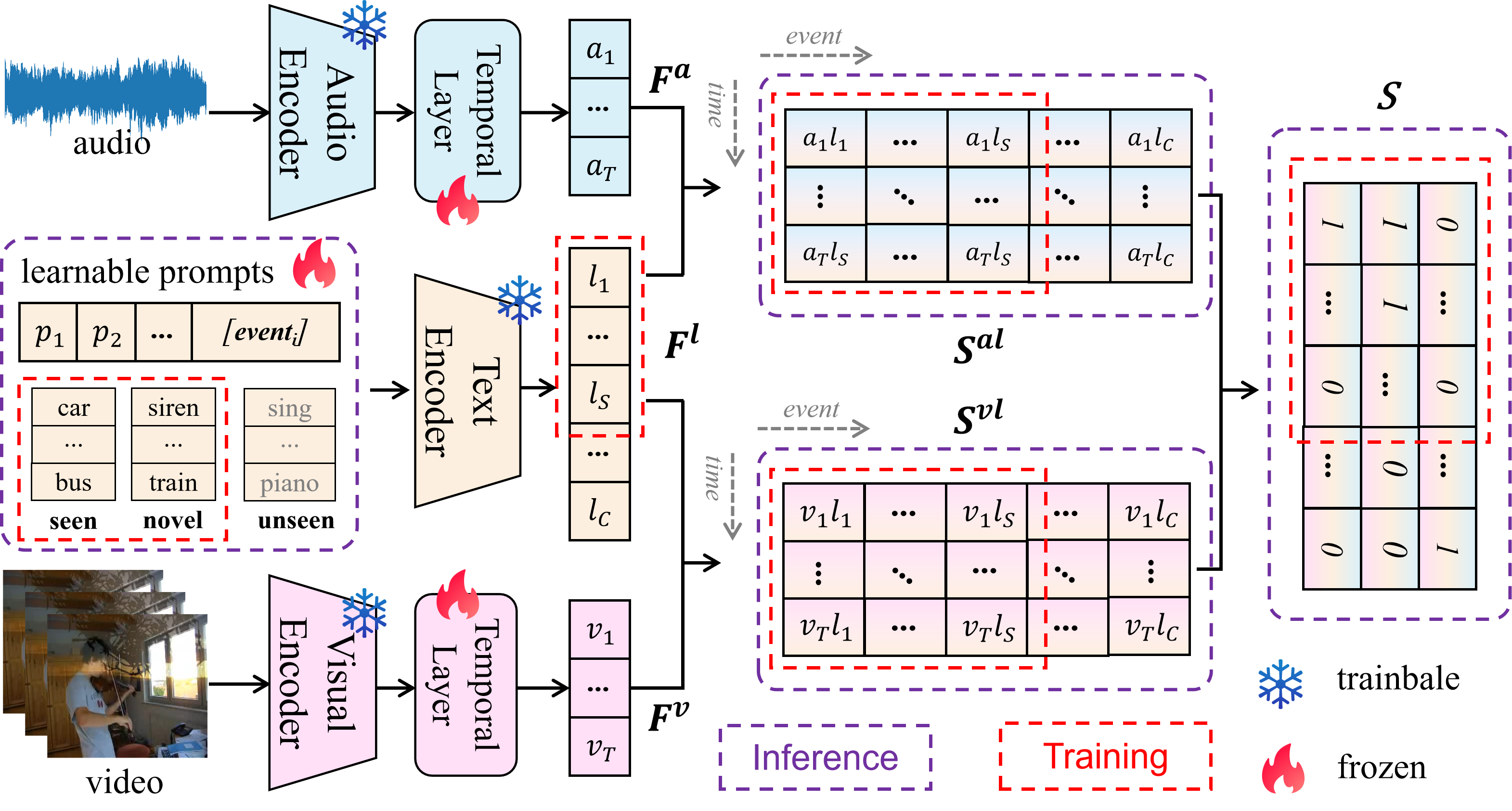}
    \caption{Overview of the proposed \textbf{S}emantic Structure-al\textbf{I}gned Pro\textbf{MP}t \textbf{LE}arning (SIMPLE) framework for open-vocabulary AVEL.
    }
    \label{fig:framework}
\end{figure}

The set of learnable prompt embeddings for all seen event categories is collectively defined as $\textbf{\textit{F}}^l = \{ \textbf{\textit{l}}_i \}_{i=1}^{C^s}$,
which serves  as semantic prototypes of audio-visual events 
to facilitate cross-modal matching.
We then compute pairwise cosine similarities 
between  segment-level audio-visual features and learnable prompt embeddings
to construct two modality-specific similarity matrices
$\textbf{\textit{S}}^{al} \in \mathbb{R}^{T \times C^s}$ and
$\textbf{\textit{S}}^{vl} \in \mathbb{R}^{T \times C^s}$:
\begin{equation}
\textbf{\textit{S}}^{al} = \| \textbf{\textit{F}}^a \| \otimes \| \textbf{\textit{F}}^l \|^T,
\quad
\textbf{\textit{S}}^{vl} = \| \textbf{\textit{F}}^v \| \otimes \| \textbf{\textit{F}}^l \|^T,
\label{Eq4}
\end{equation}
where $\| \cdot \|$ denotes row-wise L2-normalization, $\otimes$ indicates  matrix multiplication,
and $(\cdot)^T$ indicates matrix transposition.
The final prediction $\textbf{\textit{S}}$ is obtained by fusing these two  matrices from Eq.~\ref{Eq4}
via equal-weight averaging.
We minimize  the cross-entropy loss 
$\mathcal{L}_{ce}$
between $\textbf{\textit{S}}$ and $\textbf{\textit{Y}}^s$ over the set of seen event  samples.

\begin{figure*}[!t] 
    \centering \includegraphics[width=0.99\linewidth]{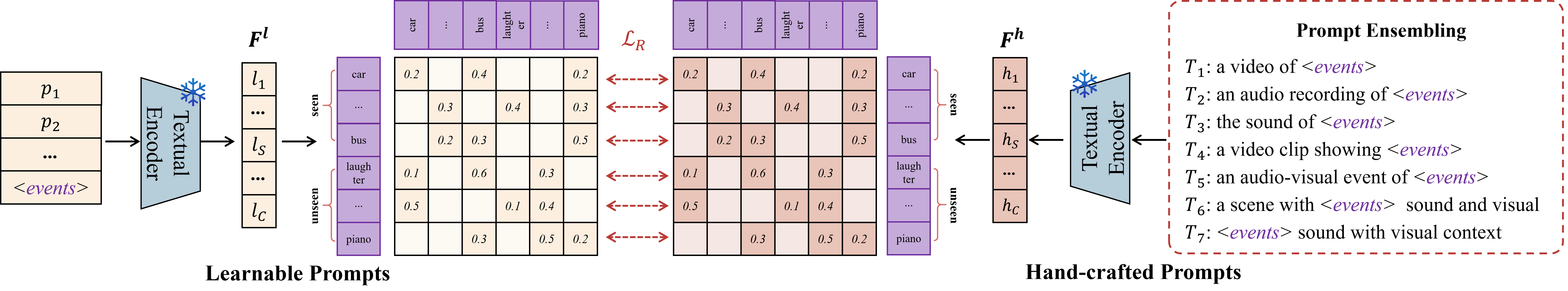}
    \caption{Overview of Semantic Topology Consistency Regularization (STCR) within Learnable Continuous Prompt Tuning (LCPT). 
    Our STCR operates on both seen and synthetic novel categories,
    aligning the pairwise semantic relationships among text embeddings generated by learnable prompts with those derived from an ensemble of hand-crafted templates. }
    \label{fig5}
\end{figure*}

However,
since our LCPT is trained exclusively on seen event categories,
the learnable prompts formulated in Eq.~\ref{Eq3} 
are prone to overfitting to the semantic distribution of seen categories.
As a result,
the native inter-event semantic topology encoded within pre-trained text encoder $\boldsymbol{\phi}_{txt}$ is distorted at inference time.
This phenomenon,
illustrated in Fig.~\ref{fig:drift},
corresponds to semantic topology drift.
Once the learnable prompts produce distorted pairwise semantic relationships across categories, 
their discriminative capacity degrades, leading to erroneous predictions.
Further mechanism analyses and discussions are provided in  Sect.~\ref{Exper} and \supplementary{Supplementary Materials}.

\subsubsection{Semantic Topology Consistency Regularization with Random Category Sampling}
\label{sect:STCR}

To mitigate semantic topology drift,
we introduce the STCR mechanism to regularize 
the  embeddings $\boldsymbol{l}_i$ from  Eq.~\ref{Eq3},
preserving inter‑event semantic relationships.
Concretely,
we feed the full list of seen event categories into Qwen2-7B \cite{Yang2024Qwen2TR} to generate $C^n$  synthetic novel categories 
whose semantics are consistent with the seen categories.
These categories serve as simulated unseen event categories throughout training.
Critically,
these synthetic categories are generated entirely without access to any test-set information and strictly verified to be fully disjoint from
the benchmark’s real unseen categories, eliminating all risk
of test-set information leakage.
The complete set of synthetic novel categories  is provided in the \supplementary{Supplementary Materials}.

To obtain more robust and  representative embeddings per event category,
we ensemble text embeddings derived from $T$ hand-crafted templates  following Eq.~\ref{Eq2}.
Let $\textbf{\textit{h}}_i^j$
be the $j$-th hand‑crafted template for the $i$-th event category.
The ensemble  embedding for the $i$-th event category
is defined as
$\textbf{\textit{h}}_i = \frac{1}{T} \sum_{j=1}^T \textbf{\textit{h}}_i^j$.
We collect the ensemble  embeddings of all seen and novel categories into the set 
$\textbf{\textit{F}}^h = \{ \textbf{\textit{h}}_i \}_{i=1}^{M}$,
where $M=C^s+C^n$.
Similarly,
we define $\textbf{\textit{F}}^l = \{ \textbf{\textit{l}}_i \}_{i=1}^{M}$
as the corresponding embeddings generated by learnable prompts through Eq.\ref{Eq3}.

A straightforward strategy for semantic topology regularization 
is to construct full pairwise similarity matrices 
for $\textbf{\textit{F}}^h$ and $\textbf{\textit{F}}^l$,
then minimize the global KL divergence
across all inter‑category pairs. 
Nevertheless,
such a rigid global constraint 
tends to cause severe overfitting to seen categories. 
To address this issue,
STCR follows \cite{jung2025learning} 
and samples a random subset of event categories,
which injects stochasticity into the regularization process.
In detail,
for each event category $i$,
we uniformly sample $K$ distinct events without replacement
from the remaining $M-1$ events  
to form the subset $\mathcal{I}_i$.
The collection of all sampled subsets is formalized as:
\begin{equation}
\begin{aligned}
    \mathcal{I} = \{\mathcal{I}_1,\mathcal{I}_2,\dots,\mathcal{I}_M\}, \\
    \text{s.t.}~~ |\mathcal{I}_i| = K~~ \text{and} ~~i \notin \mathcal{I}_i,
\end{aligned}
\label{Eq6}
\end{equation}
where $i \notin \mathcal{I}_i$ excludes  self-similarity computation.

For the ensemble embeddings $\textbf{\textit{F}}^h$,
the local similarity distribution matrix $\textbf{\textit{C}}_{\mathcal{I}}^h$
is formulated as:
\begin{equation}
\textbf{\textit{C}}_{\mathcal{I}}^h
=
\begin{bmatrix}
\sigma([\cos(\textbf{\textit{h}}_1,\textbf{\textit{h}}_k)
\mid k \in \mathcal{I}_{1}])\\
\sigma([\cos(\textbf{\textit{h}}_2,\textbf{\textit{h}}_k)
\mid k \in \mathcal{I}_{2}])\\
\vdots\\
\sigma([\cos(\textbf{\textit{h}}_M,\textbf{\textit{h}}_k)
\mid k \in \mathcal{I}_{M}])
\end{bmatrix}
\in \mathbb{R}^{M\times K},
\label{Eq7}
\end{equation}
where $\sigma(\cdot)$ denotes the softmax activation function.
Similarly,
we derive the local similarity distribution matrix $\textbf{\textit{C}}_{\mathcal{I}}^l \in \mathbb{R}^{M\times K}$ 
for learnable  embeddings $\textbf{\textit{F}}^l$
using the same sampled subset $\mathcal{I}$.
The proposed STCR regularizer minimizes the row-wise KL divergence 
between  the two local similarity matrices:
\begin{equation}
\mathcal{L}_{R} = \frac{1}{M}
\sum_{i=1}^M \mathcal{KL}
\big(\textbf{\textit{C}}_{\mathcal{I},i}^h \,\|\, \textbf{\textit{C}}_{\mathcal{I},i}^l\big).
\label{Eq8}
\end{equation}
Notably, the subset $\mathcal{I}$ in Eq.~\ref{Eq6} is dynamically resampled at each parameter update step during training.

Compared with the global topology constraint,
our random category sampling strategy delivers two core advantages.
First, resampling the category subset $\mathcal{I}$  prevents learnable prompts from overfitting to a fixed semantic topology.
Second, performing distribution matching with only $K$ sampled categories per event 
reduces the computational complexity from $\mathcal{O}(M^{2})$ to $\mathcal{O}(MK)$, where $K \ll M$,
substantially reducing the training overhead.

\subsection{Training and Inference}
The overall training objective of our SIMPLE framework is formulated as:
\begin{equation}
 \mathcal{L} = \mathcal{L}_{ce} + \lambda \mathcal{L}_{R}
\label{Eq9}
\end{equation}
where $\lambda$ is a hyperparameter balancing the two loss terms.

During inference,
the open-vocabulary AVEL requires  processing audio-visual samples 
containing  both seen and unseen event categories.
As illustrated in the purple dashed box in Fig.\ref{fig:framework},
we replace the category word embedding $\boldsymbol{c}_i$ in Eq.\ref{Eq3},
with embeddings of all seen and unseen events from the target dataset
while all $P$ context vectors remain frozen.
The forward propagation pipeline for audio and visual modalities remains identical to that in the training stage.

\begin{table*}[t]
\centering
\caption{Performance comparison with state-of-the-art (SOTA) methods on OV-AVEBench.
Bold and underlined values denote the best and second-best results, respectively.}
\begin{tabular}{c cccc cccc cccc}
\toprule
\multirow{2}{*}[-1ex]{\textbf{Method}} & 
\multicolumn{4}{c}{\textbf{Seen}} & 
\multicolumn{4}{c}{\textbf{Unseen}} & 
\multicolumn{4}{c}{\textbf{Total}} \\

\cmidrule(lr){2-5} \cmidrule(lr){6-9} \cmidrule(lr){10-13}

& \textit{Acc}. & \textit{Seg}. & \textit{Eve}. & \textit{Avg}. & \textit{Acc}. & \textit{Seg}. & \textit{Eve}. & \textit{Avg}. & \textit{Acc}. & \textit{Seg}. & \textit{Eve}. & \textit{Avg}. \\
\midrule
Video-LLaMA2\cite{cheng2024videollama} & 50.1 & 40.6 & 32.0 & 40.9 & 48.5 & 38.5 & 29.0 & 38.6 & 48.9 & 39.1 & 29.8 & 39.3 \\

CLIP\cite{radford2021learning} \& CLAP \cite{wu2023large} & 51.4 & 41.4 & 31.9 & 41.6 & 51.6 & 42.2 & 31.6 & 41.8 & 51.5 & 41.9 & 31.7 & 41.7 \\

OVAVE-FT \cite{zhou2025towards} & 57.5 & 45.0 & 34.0 & 45.5 & 59.8 & 47.3 & 34.0 & 47.0 & 59.2 & 46.7 & 34.0 & 46.6 \\
\midrule

CMRA \cite{xu2020cross} & 65.2 & 58.8 & 54.3 & 59.4 & 36.0 & 31.0 & 26.3 & 31.1 & 44.3 & 38.9 & 34.3 & 39.2 \\

AVE \cite{tian2018audio} & \underline{76.6} & 63.6 & 56.0 & 65.4 & 44.6 & 33.2 & 24.0 & 34.0 & 53.8 & 41.9 & 33.2 & 42.9 \\

PSP \cite{zhou2021positive} & 75.4 & 66.8 & 61.0 & 67.7 & 33.7 & 28.2 & 24.2 & 28.7 & 45.6 & 39.3 & 34.7 & 39.9 \\

MM-Pyramid \cite{yu2022mm} & 76.5 & \underline{66.9} & 62.3 & 68.6 & 36.8 & 29.0 & 23.8 & 29.9 & 48.4 & 40.2 & 35.2 & 41.2 \\

OVAVE-FT \cite{zhou2025towards} & 72.5 & 61.8 & 54.5 & 62.9 & 64.9 & 55.0 & 47.5 & 55.8 & 67.1 & 56.9 & 49.5 & 57.8 \\

HSCHG\cite{yang2026hierarchical} & 76.4 & 66.8 & \underline{63.1} & \underline{68.8} & 65.6 & 55.7 & 48.2 & 56.5 & 68.5 & 58.8 & 51.6 & \underline{59.7}\\

$\text{AV}^2$A-Mual\cite{jocompositional} & 53.0 & 46.1 & 42.7 & 47.3 & 59.5 & 51.1 & 46.4 & 52.3 & 57.6 & 49.7 & 45.4 & 50.9\\

$\text{AV}^2$A-Grid\cite{jocompositional} & 59.9 & 53.1 & 49.1 & 54.1 & 63.4 & 55.6 & 50.9 & 56.6 & 62.4 & 54.9 & 50.4 & 55.9\\

Cue2Rule\cite{jocompositional} & 68.6 & 60.1 & 54.6 & 61.1 & \underline{68.5} & \underline{59.1} & \underline{52.0} & \underline{59.9} & \underline{68.6} & \textbf{59.4} & \textbf{52.7} & \textbf{60.2}\\

\rowcolor{gray!20} 
\textbf{Ours} & \textbf{79.1} & \textbf{70.0} & \textbf{64.5} & \textbf{71.2} & \textbf{71.3} & \textbf{61.5} & \textbf{52.3} & \textbf{61.7} & \textbf{69.8} & \underline{59.3} & \underline{51.4} & \textbf{60.2}\\ 

\bottomrule
\end{tabular}
\label{SOTA}
\end{table*}
\begin{table}[t]
\centering
\caption{Optimization efficiency comparison on OV-AVEBench.
``Params.'' indicates the number of updated parameters, 
while ``Time''
denotes the training time until obtaining the optimal checkpoint.}
\label{tab:efficiency}
\begin{tabular}{cccc}
\toprule
\textbf{Method}
& \textbf{Params}. $\downarrow$
& \textbf{Time} $\downarrow$
& \textbf{Optimization} \\
\midrule
 OVAVE-FT \cite{zhou2025towards}
& 8.4M
&  \(\approx\) 6.0h
& Direct training \\
Cue2Rule \cite{jocompositional}
& N/R
& 9.8 h/session
& $3\times20$ search \\
\rowcolor{gray!20}
\color{RoyalBlue} {Ours}
& \color{RoyalBlue} {4.2M}
& \color{RoyalBlue} {\(\approx\) 2.0h}
& {Direct training} \\
\bottomrule
\end{tabular}
\end{table}

\section{Experiments}
\label{Exper}

\subsection{Experimental Setting}

\subsubsection{Dataset}
We conduct comprehensive experiments on the OV‑AVEBench benchmark \cite{zhou2025towards}, 
a benchmark tailored for the open‑vocabulary AVEL task.
It removes highly similar and rarely‑occurring real‑world events from the original 309 VGGSound \cite{chen2020vggsound} event categories and retains 67 common real-world event categories. 
These events cover five major semantic groups: 
human activities, 
animal activities, 
musical instruments, 
vehicles, 
and other real‑life audio‑visual scenes. 
The dataset contains 24,800 videos in total, with 13,182 samples for training, 5,798 for validation, and 5,820 for testing.
Under the open‑vocabulary setting, 
46 out of the 67 categories serve as seen event categories available during training, 
while the remaining 21 unseen categories appear only in the validation and testing phases. 
To rigorously evaluate the model’s generalization capability toward unseen event categories, the ratio of videos containing seen events to those containing unseen events is kept at roughly $3:7$ within both validation and test splits.

\subsubsection{Evaluation Metrics}
To guarantee reproducibility and  fair comparison, 
we adopt the same seen/unseen split, data partitions, and evaluation protocol as the original baseline study~\cite{zhou2025towards}.
Concretely,
four metrics are used to assess performance on seen, unseen, and total test splits: segment‑level accuracy (\textit{Acc}.), segment‑level F1‑score (\textit{Seg}.), event‑level F1‑score (\textit{Eve}.), and their average (\textit{Avg}.).
\textit{Acc}. quantifies the proportion of correctly classified one-second segments.
\textit{Seg}. is the harmonic mean of precision and recall over all test segments, where each segment is treated as an independent sample.
\textit{Eve}. evaluates performance at the event instance level. 
Consecutive segments with identical predictions are merged into  event proposals. 
A prediction is marked as a true positive if its IoU with the corresponding GT event exceeds 0.5. 
Precision and recall are calculated on merged events, yielding the event‑level F1‑score.
\textit{Avg}. is the arithmetic mean of \textit{Acc}., \textit{Seg}., and \textit{Eve}., serving as a comprehensive indicator of overall performance.

\subsubsection{Implementation Details}

All extracted embeddings are  projected into a shared 1024-dimensional space.
Each modal-specific  temporal layer
uses a hidden dimension of 256, one attention head, a feed-forward network with an intermediate size of 512, a dropout rate of 0.1, and post-layer normalization.
We insert $P=16$ learnable context vectors
initialized from $\mathcal{N}(0, 0.02)$
after the start-of-sequence token.
In total, 
the number of trainable parameters is approximately 4.2M.
To enhance semantic diversity for regularization,
we incorporate 
$C^n=170$ synthetic novel event category names.
We fix the random sampling size $K$ in Eq.~\ref{Eq6} to 64
and the regularization weight $\lambda$ in Eq.~\ref{Eq9} to 0.1.

For model training, we employ the Adam optimizer with $\beta_1=0.9$ and $\beta_2=0.999$. 
The initial learning rate is set to $5\times 10^{-5}$, 
paired with a weight decay coefficient of $5\times 10^{-4}$. 
Training is conducted with a batch size of 16, and gradient clipping is applied with a maximum norm of 0.8. 
We adopt a multi-step scheduler that decays the learning rate by a factor of 0.5 at epochs 10, 20, and 30, respectively. 
The model is trained for up to 36 epochs, with early stopping triggered based on the \textit{Avg.} metric evaluated on the validation split (5,798 videos covering all 67 seen and unseen categories).
All experiments are conducted on a single NVIDIA RTX 4090 GPU with 24GB memory.

\begin{table*}[t]
\centering
\caption{Generalization performance of the proposed framework across seven diverse hand‑designed prompts ($T_1 \sim T_7$). 
Upward arrows ($\uparrow$) denote performance improvements, and downward arrows ($\downarrow$) denote performance degradation compared to the baseline model with vanilla prompts.
The detailed definitions of all prompts are listed in Fig.\ref{fig5}.
}
\begin{tabular}{c cccc cccc cccc}
\toprule
\multirow{2}{*}[-1ex]{\textbf{Prompt}} & 
\multicolumn{4}{c}{\textbf{Seen}} & 
\multicolumn{4}{c}{\textbf{Unseen}} & 
\multicolumn{4}{c}{\textbf{Total}} \\

\cmidrule(lr){2-5} \cmidrule(lr){6-9} \cmidrule(lr){10-13}

& \textit{Acc}. & \textit{Seg}. & \textit{Eve}. & \textit{Avg}. & \textit{Acc}. & \textit{Seg}. & \textit{Eve}. & \textit{Avg}. & \textit{Acc}. & \textit{Seg}. & \textit{Eve}. & \textit{Avg}. \\
\midrule


$T_1$ \cite{kurzendorfer2024audio} & 78.9  & 70.8 & 65.6 & 71.8 & 69.2 & 58.7 & 49.2 & 59.0 & 67.7 & 57.4 & 49.5 & 58.2\\
\rowcolor{gray!20} \color{RoyalBlue}
$+$ Ours & \color{RoyalBlue} 0.1 $\uparrow$ & 0.5 $\downarrow$ & 0.8 $\downarrow$ & 0.4 $\downarrow$ & \color{RoyalBlue} 1.5 $\uparrow$ &  \color{RoyalBlue} 2.0 $\uparrow$ & \color{RoyalBlue} 2.2 $\uparrow$ & \color{RoyalBlue} 1.9 $\uparrow$ & \color{RoyalBlue} 1.8 $\uparrow$ & \color{RoyalBlue} 1.8 $\uparrow$ & \color{RoyalBlue} 1.8 $\uparrow$ & \color{RoyalBlue} 1.8 $\uparrow$ \\
\midrule

$T_2$ \cite{hanif2024palm} & 78.0 & 68.9 & 63.3 & 70.1 & 68.2 & 57.6 & 48.1 & 58.0 & 65.9 & 55.0 & 46.8 & 55.9 \\
\rowcolor{gray!20} \color{RoyalBlue}
$+$ Ours & \color{RoyalBlue} 0.9 $\uparrow$ & \color{RoyalBlue} 1.1 $\uparrow$ & \color{RoyalBlue} 1.5 $\uparrow$ & \color{RoyalBlue} 1.2 $\uparrow$ & \color{RoyalBlue} 1.4 $\uparrow$ & \color{RoyalBlue} 2.2 $\uparrow$ & \color{RoyalBlue} 2.6 $\uparrow$ & \color{RoyalBlue} 2.1 $\uparrow$ & \color{RoyalBlue} 2.5
$\uparrow$ & \color{RoyalBlue} 2.8 $\uparrow$ & \color{RoyalBlue} 2.9 $\uparrow$ & \color{RoyalBlue} 2.9 $\uparrow$ \\
\midrule

$T_3$ \cite{hanif2024palm} & 78.4 & 70.4 & 65.0 & 71.3 & 68.1 & 58.5 & 49.4 & 58.7 & 67.3 & 57.3 & 49.6 & 58.1 \\
\rowcolor{gray!20} \color{RoyalBlue}
$+$ Ours & \color{RoyalBlue} 0.5 $\uparrow$ & 0.5 $\downarrow$ & 0.9 $\downarrow$ & 0.3 $\downarrow$ & \color{RoyalBlue} 1.2 $\uparrow$ & \color{RoyalBlue} 0.7 $\uparrow$ & \color{RoyalBlue} 0.2 $\uparrow$ & \color{RoyalBlue} 0.7 $\uparrow$ & \color{RoyalBlue} 1.0 $\uparrow$ & \color{RoyalBlue} 0.2 $\uparrow$ & 
0.4 $\downarrow$ & 
\color{RoyalBlue} 0.3 $\uparrow$ \\
\midrule

$T_4$ \cite{hanif2024palm} & 78.4 & 69.3 & 63.5 & 70.4 & 69.0 & 58.6 & 49.0 & 58.9 & 66.4 & 56.0 & 47.9 & 56.8 \\
\rowcolor{gray!20} \color{RoyalBlue}
$+$ Ours & \color{RoyalBlue} 0.7 $\uparrow$ & \color{RoyalBlue} 0.7 $\uparrow$ & \color{RoyalBlue} 1.0 $\uparrow$ & \color{RoyalBlue} 0.8 $\uparrow$ & \color{RoyalBlue} 2.3 $\uparrow$ & \color{RoyalBlue} 2.9 $\uparrow$ & \color{RoyalBlue} 3.3 $\uparrow$ & \color{RoyalBlue} 2.8 $\uparrow$ & \color{RoyalBlue} 3.4
$\uparrow$ & \color{RoyalBlue} 3.3
$\uparrow$ & \color{RoyalBlue} 3.5 $\uparrow$ & \color{RoyalBlue} 3.4 $\uparrow$ \\
\midrule

$T_5$\cite{zhou2025towards} & 78.5 & 70.3 & 64.7 & 71.2 & 68.6 & 60.0 & 50.0 & 59.5 & 67.0 & 57.3 & 49.5 & 57.9 \\
\rowcolor{gray!20} \color{RoyalBlue}
$+$ Ours & \color{RoyalBlue} 0.5 $\uparrow$ & 0.1 
$\downarrow$ & \color{RoyalBlue} 0.1 $\downarrow$ & \color{RoyalBlue} 0.1 $\uparrow$ & \color{RoyalBlue} 1.1 $\uparrow$ & 0.9 
$\downarrow$ & 0.8 
$\downarrow$ & 0.2 
$\downarrow$ & \color{RoyalBlue} 1.2
$\uparrow$ & \color{RoyalBlue} 0.1 $\downarrow$ & \color{RoyalBlue} 0.8 $\downarrow$ & \color{RoyalBlue} 0.1 $\uparrow$ \\
\midrule

$T_6$\cite{jung2025learning} & 78.6 & 70.2 & 64.6 & 71.2 & 66.2 & 57.0 & 47.7 & 57.0 & 66.1 & 56.5 & 48.3 & 57.0 \\
\rowcolor{gray!20} \color{RoyalBlue}
$+$ Ours & \color{RoyalBlue} 0.4 $\uparrow$ & 0.2 $\downarrow$ & 0.2 $\downarrow$ & \color{RoyalBlue} 0.0 & \color{RoyalBlue} 5.3 $\uparrow$ & \color{RoyalBlue} 4.1 $\uparrow$ & \color{RoyalBlue} 4.0 $\uparrow$ & \color{RoyalBlue} 4.5 $\uparrow$ & \color{RoyalBlue} 3.7 $\uparrow$ & \color{RoyalBlue} 2.3 $\uparrow$ & \color{RoyalBlue} 2.4 $\uparrow$ & \color{RoyalBlue} 2.8 $\uparrow$ \\
\midrule

$T_7$\cite{jung2025learning} & 78.8 & 70.5 & 64.9 & 71.4 & 67.3 & 57.6 & 48.2 & 57.7 & 66.3 & 56.5 & 48.4 & 57.1 \\
\rowcolor{gray!20} \color{RoyalBlue}
$+$ Ours & \color{RoyalBlue} 0.2 $\uparrow$ & 0.5 $\downarrow$ & 0.7 $\downarrow$ & 0.3 $\downarrow$ & \color{RoyalBlue} 3.6 $\uparrow$ & \color{RoyalBlue} 3.0 $\uparrow$ & \color{RoyalBlue} 3.1 $\uparrow$ & \color{RoyalBlue} 3.2 $\uparrow$ & \color{RoyalBlue} 3.0 $\uparrow$ & \color{RoyalBlue} 2.1 $\uparrow$ & \color{RoyalBlue} 2.0 $\uparrow$ & \color{RoyalBlue} 2.4 $\uparrow$ \\
\midrule


Ensemble & 79.0 & 70.7 & 65.7 & 71.8 & 69.6 &59.6 & 50.5 & 59.9 & 67.5 & 57.4 & 49.8 & 58.2 \\
\rowcolor{gray!20} \color{RoyalBlue}
$+$ Ours & \color{RoyalBlue} 0.0  & 0.6 $\downarrow$ & 0.9 $\downarrow$ & 0.5 $\downarrow$ & \color{RoyalBlue} 1.6 $\uparrow$ & \color{RoyalBlue} 1.7 $\uparrow$ &\color{RoyalBlue} 1.8 $\uparrow$ & \color{RoyalBlue} 1.7 $\uparrow$ & \color{RoyalBlue} 2.3 $\uparrow$ & \color{RoyalBlue} 1.8 $\uparrow$ & \color{RoyalBlue} 1.7 $\uparrow$ & \color{RoyalBlue} 1.9 $\uparrow$ \\

\bottomrule
\end{tabular}
\label{tab2}
\end{table*}

\subsection{Quantitative Results and Analysis}

\subsubsection{Comparisons with SOTAs}
Tab.~\ref{SOTA} summarizes the comprehensive performance comparison
between the proposed approach and  12 SOTA methods on the OV-AVEBench benchmark for open-vocabulary AVEL.
Our method achieves the best performance across all four evaluation
metrics on both the seen and unseen test splits, and attains an average score of 60.2\% on the total split, 
matching the previous best-performing Cue2Rule
\cite{jocompositional}.
More importantly, our method substantially improves the performance
on both seen and unseen categories, with average scores increasing
from 68.8\% to 71.2\% and from 59.9\% to 61.7\%, respectively.
This corresponds to absolute improvements of
$+2.4$~\textit{pp} and $+1.8$~\textit{pp}
over the best prior results.

General-purpose baselines, including Video-LLaMA2~\cite{cheng2024videollama},
CLIP~\cite{radford2021learning}, CLAP~\cite{wu2023large}, and
OVAVE-FT~\cite{zhou2025towards}, achieve only 39.3\%--46.6\% on the total Avg. metric.
Without task-specific adaptation, pre-trained cross-modal features
struggle to support fine-grained AVEL.
While closed-set AVEL methods, including CMRA~\cite{xu2020cross},
AVE~\cite{tian2018audio}, PSP~\cite{zhou2021positive}, and
MM-Pyramid~\cite{yu2022mm}, achieve strong performance on seen
categories (Avg. 59.4\%--68.6\%), their performance declines sharply
for unseen categories (Avg. 28.7\%--34.0\%).
Notably, AVE~\cite{tian2018audio} exhibits a large seen--unseen
Avg. gap of 31.4~\textit{pp}, indicating severe overfitting to
seen-category semantics.

Existing open-vocabulary AVEL methods typically encounter the inherent
trade-off between seen-category classification and unseen-category
generalization.
OVAVE-FT~\cite{zhou2025towards} improves unseen-split performance to
55.8\% but only achieves 62.9\% on the seen split.
HSCHG~\cite{yang2026hierarchical} reaches 68.8\% on seen categories,
while maintaining an unseen-split Avg. of 56.5\%.
The AV$^2$A variants~\cite{jocompositional} improve unseen-split
generalization, but at the expense of substantially significantly degraded seen-split performance.
For example,
AV$^2$A-Mual obtains only 47.3\% on the seen
split.
Cue2Rule~\cite{jocompositional} provides a more balanced performance
between seen and unseen categories and achieves a Total Avg. of 60.2\%.
Unlike these existing approaches, our method achieves substantially
stronger performance on both seen and unseen splits, while matching
Cue2Rule on the overall Total Avg. metric.

\subsubsection{Efficiency and Complexity Analysis}

We further evaluate the parameter efficiency and optimization overhead of our method, 
as summarized in Tab.~\ref{tab:efficiency}. 
Our method achieves
a total Avg. of $60.2$ while updating only $4.2$M trainable parameters, 
in comparison to $8.4$M required by the OVAVE-FT~\cite{zhou2025towards}. 
This amounts to roughly a $50\%$ reduction in trainable parameters,
accompanied by a performance gain 
where the total Avg. improves
from $57.8\%$ to $60.2\%$. 
Beyond parameter savings, our best checkpoint is obtained in approximately 2 hours of training.

Compared with Cue2Rule~\cite{jocompositional}, 
our method matches its SOTA
total Avg. of $60.2\%$ but avoids iterative agentic rule search. 
In contrast, Cue2Rule takes 9.8 hours per search session and runs three separate sessions in total. 
Furthermore,
our STCR mechanism is activated only at training and
introduces no additional computational overhead during inference. 
These results demonstrate that our method achieves new SOTA performance while substantially cutting both parameter and optimization costs.

\subsubsection{Generalization across diverse hand-designed prompts}
Tab.~\ref{tab2} assesses the generalization capability of our framework across seven diverse hand-designed prompts ($T_1$--$T_7$).
Consistent performance gains are observed across the seen, unseen, and total splits.
In particular, with the visual-oriented prompt $T_4$~\cite{hanif2024palm}, our method yields the largest gain on the total split in terms of the Avg. metric (+3.4 \textit{pp}), which is the most important metric in our evaluation.
Moreover, when averaged across all seven prompts, our method improves the total Avg. from 58.2\% to 60.2\%, corresponding to a gain of 2.0 \textit{pp}.
This further confirms the robustness of our method to variations in hand-crafted prompts.
Whether a prompt prioritizes visual cues ($T_1$~\cite{kurzendorfer2024audio} and $T_4$~\cite{hanif2024palm}), audio cues ($T_2$~\cite{hanif2024palm} and $T_3$~\cite{hanif2024palm}), or both modalities ($T_5$~\cite{zhou2025towards}), our method achieves consistent performance gains, demonstrating robustness and broad applicability.

\begin{table*}[t]
\centering
\caption{Validity verification about different ratios of training samples compared with OVAVE-FT \cite{zhou2025towards}.}
\begin{tabular}{cc cccc cccc cccc}
\toprule
\multirow{2}{*}[-1ex]{\textbf{Training}} & 
\multirow{2}{*}[-1ex]{\textbf{Method}} & 
\multicolumn{4}{c}{\textbf{Seen}} & 
\multicolumn{4}{c}{\textbf{Unseen}} & 
\multicolumn{4}{c}{\textbf{Total}} \\

\cmidrule(lr){3-6} \cmidrule(lr){7-10} \cmidrule(lr){11-14}

& & \textit{Acc}. & \textit{Seg}. & \textit{Eve}. & \textit{Avg}. & \textit{Acc}. & \textit{Seg}. & \textit{Eve}. & \textit{Avg}. & \textit{Acc}. & \textit{Seg}. & \textit{Eve}. & \textit{Avg}. \\
\midrule

\multirow{2}{*}{25\%}  & OVAVE-FT \cite{zhou2025towards} & 73.0  & 62.6  & 55.7  & 63.8 & 63.8  & 54.4  & 47.5  & 55.2 & 66.4  & 56.8  & 49.8  & 57.7 \\
& \cellcolor{gray!20}  \color{RoyalBlue} Ours & \cellcolor{gray!20}  \color{RoyalBlue}76.1 & \cellcolor{gray!20} \color{RoyalBlue}66.4 & \cellcolor{gray!20} \color{RoyalBlue}60.4 & \cellcolor{gray!20} \color{RoyalBlue}67.6 & \cellcolor{gray!20} \color{RoyalBlue}71.7 & \cellcolor{gray!20} \color{RoyalBlue}60.7 & \cellcolor{gray!20} \color{RoyalBlue}51.3 & \cellcolor{gray!20} \color{RoyalBlue}61.2 & \cellcolor{gray!20} \color{RoyalBlue}68.9 & \cellcolor{gray!20} \color{RoyalBlue}57.3 & \cellcolor{gray!20} \color{RoyalBlue}49.1 &  \cellcolor{gray!20} \color{RoyalBlue}58.4 \\
\midrule

\multirow{2}{*}{50\%}  & OVAVE-FT \cite{zhou2025towards} & 75.3  & 66.0  & 59.9  & 67.1 & 63.5  &  53.5 & 45.6  & 54.2  & 66.7  & 57.1  & 49.7  & 57.8 \\
& \cellcolor{gray!20} \color{RoyalBlue} Ours & \cellcolor{gray!20} \color{RoyalBlue} 77.4 & \cellcolor{gray!20} \color{RoyalBlue} 68.2 & \cellcolor{gray!20} \color{RoyalBlue} 62.5 & \cellcolor{gray!20} \color{RoyalBlue} 69.4 & \cellcolor{gray!20} \color{RoyalBlue} 71.9 & \cellcolor{gray!20} \color{RoyalBlue} 61.3 & \cellcolor{gray!20} \color{RoyalBlue} 52.0 & \cellcolor{gray!20} \color{RoyalBlue} 61.7 & \cellcolor{gray!20} \color{RoyalBlue} 69.6 & \cellcolor{gray!20} \color{RoyalBlue} 58.5 & \cellcolor{gray!20} \color{RoyalBlue} 50.5 &  \cellcolor{gray!20} \color{RoyalBlue} 59.5 \\
\midrule

\multirow{2}{*}{75\%}  & OVAVE-FT \cite{zhou2025towards} & 75.1  & 65.4  & 59.3  & 66.6 & 62.5  & 53.5  & 46.1  & 54.0 & 66.1  & 56.9  &  49.9 & 57.6
\\
& \cellcolor{gray!20} \color{RoyalBlue} Ours & \cellcolor{gray!20} \color{RoyalBlue} 78.3 & \cellcolor{gray!20} \color{RoyalBlue} 69.8 & \cellcolor{gray!20} \color{RoyalBlue} 64.3 & \cellcolor{gray!20} \color{RoyalBlue} 70.8 & \cellcolor{gray!20} \color{RoyalBlue} 71.5 & \cellcolor{gray!20} \color{RoyalBlue} 61.3 & \cellcolor{gray!20} \color{RoyalBlue} 51.7 & \cellcolor{gray!20} \color{RoyalBlue} 61.5 & \cellcolor{gray!20} \color{RoyalBlue} 69.4 & \cellcolor{gray!20} \color{RoyalBlue} 58.7 & \cellcolor{gray!20} \color{RoyalBlue} 50.3&  \cellcolor{gray!20} \color{RoyalBlue} 59.4 \\
\midrule

\multirow{2}{*}{100\%} & OVAVE-FT \cite{zhou2025towards}  & 72.5  & 61.8  & 54.5  & 62.9 & 64.9  & 55.0  & 47.5  & 55.8 & 67.1  & 56.9  & 49.5  & 57.8  \\
& \cellcolor{gray!20} \color{RoyalBlue} Ours & \cellcolor{gray!20} \color{RoyalBlue} 79.1 & \cellcolor{gray!20} \color{RoyalBlue} 70.0 & \cellcolor{gray!20} \color{RoyalBlue} 64.5 & \cellcolor{gray!20} \color{RoyalBlue} 71.2 & \cellcolor{gray!20} \color{RoyalBlue} 71.3 & \cellcolor{gray!20} \color{RoyalBlue} 61.5 & \cellcolor{gray!20} \color{RoyalBlue} 52.3 & \cellcolor{gray!20} \color{RoyalBlue} 61.7 & \cellcolor{gray!20} \color{RoyalBlue} 69.8 & \cellcolor{gray!20} \color{RoyalBlue} 59.3 & \cellcolor{gray!20} \color{RoyalBlue} 51.4 &  \cellcolor{gray!20} \color{RoyalBlue} 60.2 \\

\bottomrule
\end{tabular}
\label{tab3}
\end{table*}
\begin{table*}[t]
\centering
\caption{Comparison with OVAVE-FT \cite{zhou2025towards} under different seen-to-unseen event-category ratios, where $C^s$ and $C^u$ denote the numbers of seen and unseen event categories, respectively.}
\begin{tabular}{cccc cccc cccc cccc}
\toprule
\multirow{2}{*}{$C^s$} &
\multirow{2}{*}{$C^u$} &
\multirow{2}{*}{\textbf{Ratio}} &
\multirow{2}{*}{\textbf{Method}} &
\multicolumn{4}{c}{\textbf{Seen}} &
\multicolumn{4}{c}{\textbf{Unseen}} &
\multicolumn{4}{c}{\textbf{Total}} \\
\cmidrule(lr){5-8} \cmidrule(lr){9-12} \cmidrule(lr){13-16}
 & & & & \textit{Acc}. & \textit{Seg}. & \textit{Eve}. & \textit{Avg}. & \textit{Acc}. & \textit{Seg}. & \textit{Eve}. & \textit{Avg}. & \textit{Acc}. & \textit{Seg}. & \textit{Eve}. & \textit{Avg}. \\
\midrule

\multirow{4}{*}{15} & \multirow{2}{*}{21} & \multirow{2}{*}{0.71} & OVAVE-FT \cite{zhou2025towards} & 71.0 & 59.1 & 52.4 & 60.8 & 53.5 & 39.7 & 30.7 & 41.3 & 57.1 & 43.7 & 35.1 & 45.3 \\
& & & \cellcolor{gray!20}\textcolor{RoyalBlue} {Ours} & \cellcolor{gray!20}\color{RoyalBlue} 80.0 & \cellcolor{gray!20}\color{RoyalBlue} 72.1 & \cellcolor{gray!20}\color{RoyalBlue} 67.1 & \cellcolor{gray!20}\color{RoyalBlue} 73.1 & \cellcolor{gray!20}\color{RoyalBlue} 58.9 & \cellcolor{gray!20}\color{RoyalBlue} 48.0 & \cellcolor{gray!20}\color{RoyalBlue} 38.8 & \cellcolor{gray!20}\color{RoyalBlue} 48.6 & \cellcolor{gray!20}\color{RoyalBlue} 65.2 & \cellcolor{gray!20}\color{RoyalBlue} 53.6 & \cellcolor{gray!20}\color{RoyalBlue} 44.8 & \cellcolor{gray!20}\color{RoyalBlue} 54.5 \\
& \multirow{2}{*}{52} & \multirow{2}{*}{0.29} & OVAVE-FT \cite{zhou2025towards} & 71.0 & 59.1 & 52.4 & 60.8 & 60.2 & 45.0 & 35.9 & 47.0 & 61.3 & 46.4 & 37.5 & 48.4 \\
& & & \cellcolor{gray!20}\textcolor{RoyalBlue}{Ours} & \cellcolor{gray!20}\color{RoyalBlue} 80.0 & \cellcolor{gray!20}\color{RoyalBlue} 72.1 & \cellcolor{gray!20}\color{RoyalBlue} 67.1 & \cellcolor{gray!20}\color{RoyalBlue} 73.1 & \cellcolor{gray!20}\color{RoyalBlue} 62.6 & \cellcolor{gray!20}\color{RoyalBlue} 49.3 & \cellcolor{gray!20}\color{RoyalBlue} 40.0 & \cellcolor{gray!20}\color{RoyalBlue} 50.6 & \cellcolor{gray!20}\color{RoyalBlue} 62.5 & \cellcolor{gray!20}\color{RoyalBlue} 49.6 & \cellcolor{gray!20}\color{RoyalBlue} 40.7 & \cellcolor{gray!20}\color{RoyalBlue} 50.9 \\
\midrule

\multirow{4}{*}{25} & \multirow{2}{*}{21} & \multirow{2}{*}{1.19} & OVAVE-FT \cite{zhou2025towards} & 73.7 & 62.9 & 56.4 & 64.3 & 63.6 & 51.5 & 42.9 & 52.7 & 66.3 & 54.6 & 46.6 & 55.8 \\
& & & \cellcolor{gray!20}\textcolor{RoyalBlue}{Ours} & \cellcolor{gray!20}\color{RoyalBlue} 80.4 & \cellcolor{gray!20}\color{RoyalBlue} 73.4 & \cellcolor{gray!20}\color{RoyalBlue} 68.9 & \cellcolor{gray!20}\color{RoyalBlue} 74.2 & \cellcolor{gray!20}\color{RoyalBlue} 70.3 & \cellcolor{gray!20}\color{RoyalBlue} 61.0 & \cellcolor{gray!20}\color{RoyalBlue} 52.9 & \cellcolor{gray!20}\color{RoyalBlue} 61.4 & \cellcolor{gray!20}\color{RoyalBlue} 64.5 & \cellcolor{gray!20}\color{RoyalBlue} 54.8 & \cellcolor{gray!20}\color{RoyalBlue} 48.4 & \cellcolor{gray!20}\color{RoyalBlue} 55.9 \\
& \multirow{2}{*}{42} & \multirow{2}{*}{0.60} & OVAVE-FT \cite{zhou2025towards} & 73.7 & 62.9 & 56.4 & 64.3 & 60.3 & 46.9 & 37.9 & 48.4 & 62.5 & 49.6 & 40.9 & 51.0 \\
& & & \cellcolor{gray!20}\textcolor{RoyalBlue}{Ours} & \cellcolor{gray!20}\color{RoyalBlue} 80.4 & \cellcolor{gray!20}\color{RoyalBlue} 73.4 & \cellcolor{gray!20}\color{RoyalBlue} 68.9 & \cellcolor{gray!20}\color{RoyalBlue} 74.2 & \cellcolor{gray!20}\color{RoyalBlue} 66.6 & \cellcolor{gray!20}\color{RoyalBlue} 55.4 & \cellcolor{gray!20}\color{RoyalBlue} 46.7 & \cellcolor{gray!20}\color{RoyalBlue} 56.2 & \cellcolor{gray!20}\color{RoyalBlue} 64.3 & \cellcolor{gray!20}\color{RoyalBlue} 53.5 & \cellcolor{gray!20}\color{RoyalBlue} 46.1 & \cellcolor{gray!20}\color{RoyalBlue} 54.6 \\
\midrule

\multirow{4}{*}{35} & \multirow{2}{*}{21} & \multirow{2}{*}{1.67} & OVAVE-FT \cite{zhou2025towards} & 73.8 & 63.3 & 56.1 & 64.4 & 63.2 & 49.6 & 40.9 & 51.2 & 66.4 & 53.7 & 45.5 & 55.2 \\
& & & \cellcolor{gray!20}\textcolor{RoyalBlue}{Ours} & \cellcolor{gray!20}\color{RoyalBlue} 76.4 & \cellcolor{gray!20}\color{RoyalBlue} 67.5 & \cellcolor{gray!20}\color{RoyalBlue} 61.1 & \cellcolor{gray!20}\color{RoyalBlue} 68.3 & \cellcolor{gray!20}\color{RoyalBlue} 70.6 & \cellcolor{gray!20}\color{RoyalBlue} 59.1 & \cellcolor{gray!20}\color{RoyalBlue} 49.7 & \cellcolor{gray!20}\color{RoyalBlue} 59.8 & \cellcolor{gray!20}\color{RoyalBlue} 66.1 & \cellcolor{gray!20}\color{RoyalBlue} 54.3 & \cellcolor{gray!20}\color{RoyalBlue} 46.1 & \cellcolor{gray!20}\color{RoyalBlue} 55.5 \\
& \multirow{2}{*}{32} & \multirow{2}{*}{1.09} & OVAVE-FT \cite{zhou2025towards} & 73.8 & 63.3 & 56.1 & 64.4 & 63.0 & 50.2 & 41.3 & 51.5 & 65.5 & 53.1 & 44.6 & 54.4 \\
& & & \cellcolor{gray!20}\textcolor{RoyalBlue}{Ours} & \cellcolor{gray!20}\color{RoyalBlue} 76.4 & \cellcolor{gray!20}\color{RoyalBlue} 67.5 & \cellcolor{gray!20}\color{RoyalBlue} 61.1 & \cellcolor{gray!20}\color{RoyalBlue} 68.3 & \cellcolor{gray!20}\color{RoyalBlue} 69.0 & \cellcolor{gray!20}\color{RoyalBlue} 57.7 & \cellcolor{gray!20}\color{RoyalBlue} 48.4 & \cellcolor{gray!20}\color{RoyalBlue} 58.4 & \cellcolor{gray!20}\color{RoyalBlue} 65.6 & \cellcolor{gray!20}\color{RoyalBlue} 54.0 & \cellcolor{gray!20}\color{RoyalBlue} 45.6 & \cellcolor{gray!20}\color{RoyalBlue} 55.1 \\
\midrule

\multirow{2}{*}{46} & \multirow{2}{*}{21} & \multirow{2}{*}{2.19} & OVAVE-FT \cite{zhou2025towards} & 72.5 & 61.8 & 54.5 & 62.9 & 64.9 & 55.0 & 47.5 & 55.8 & 67.1 & 56.9 & 49.5 & 57.8 \\
& & & \cellcolor{gray!20}\textcolor{RoyalBlue}{Ours} & \cellcolor{gray!20}\color{RoyalBlue} 79.1 & \cellcolor{gray!20}\color{RoyalBlue} 70.0 & \cellcolor{gray!20}\color{RoyalBlue} 64.5 & \cellcolor{gray!20}\color{RoyalBlue} 71.2 & \cellcolor{gray!20}\color{RoyalBlue} 71.3 & \cellcolor{gray!20}\color{RoyalBlue} 61.5 & \cellcolor{gray!20}\color{RoyalBlue} 52.3 & \cellcolor{gray!20}\color{RoyalBlue} 61.7 & \cellcolor{gray!20}\color{RoyalBlue} 69.8 & \cellcolor{gray!20}\color{RoyalBlue} 59.3 & \cellcolor{gray!20}\color{RoyalBlue} 51.4 & \cellcolor{gray!20}\color{RoyalBlue} 60.2 \\

\bottomrule
\end{tabular}
\label{tab4}
\end{table*}

\subsection{Validity Verification}

\subsubsection{Training sample numbers}

We validate the effectiveness of our framework by evaluating its
performance with different training sample sizes
(25\%, 50\%, 75\%, and 100\% of the full training set).
As shown in Tab.~\ref{tab3},
increasing the training sample size from 25\% to 100\% consistently
improves the total average score of our method from 58.4\% to 60.2\%,
with only a minor fluctuation at the 75\% level.
This trend indicates that our framework can effectively
benefit from additional training data.
Even when trained with only 25\% of the training samples,
our model achieves average scores of 67.6\%, 61.2\%, and 58.4\%
on the seen, unseen, and total splits, outperforming
OVAVE-FT~\cite{zhou2025towards} by 3.8, 6.0, and 0.7 \textit{pp},
respectively.
Remarkably, the result obtained using merely 25\% of the training
samples (58.4\%) already surpasses the full-sample result of OVAVE-FT
(57.8\%), demonstrating the superior data efficiency of our
method.

\subsubsection{Seen category numbers}

To further verify the generalization robustness of our framework, we evaluate its
performance with varying numbers of seen categories 
and compare it with OVAVE-FT~\cite{zhou2025towards}, as shown in Tab.~\ref{tab4}.
With the number of unseen categories fixed at 21, reducing the number
of seen categories from 46 to 35, 25, and 15 generally degrades the
performance of both methods, with a slight fluctuation at 25 seen
categories.
This trend aligns with the intuition that open-vocabulary generalization becomes more challenging with fewer seen categories.
Our total average drops from 60.2\% to 55.5\%, 55.9\%, and
54.5\% as the number of seen categories is reduced from 46 to 35,
25, and 15, respectively.
In comparison, OVAVE-FT~\cite{zhou2025towards} experiences a steeper decline from 57.8\%
to 55.2\%, 55.8\%, and 45.3\%.
Notably, with only 15 seen categories, our method achieves substantial
gains over OVAVE-FT across the seen, unseen, and total splits, with
improvements of 12.3, 7.3, and 9.2 \textit{pp}, respectively.
Furthermore, in the more challenging setting with 15 seen and 52
unseen categories (a seen-to-unseen ratio of  0.29), our approach still achieves a higher
total average of 50.9\%, versus 48.4\% for OVAVE-FT.
Across all configurations in Tab.~\ref{tab4}, our method consistently
outperforms OVAVE-FT in terms of the total Avg. metric.
These results demonstrate that our method retains stronger
performance as the number of seen categories decreases, indicating
greater robustness under limited seen-category supervision and
superior open-vocabulary generalization ability.

\subsection{Mechanism Analysis}

\begin{figure*}[t]
    \centering
    \includegraphics[width=\textwidth]{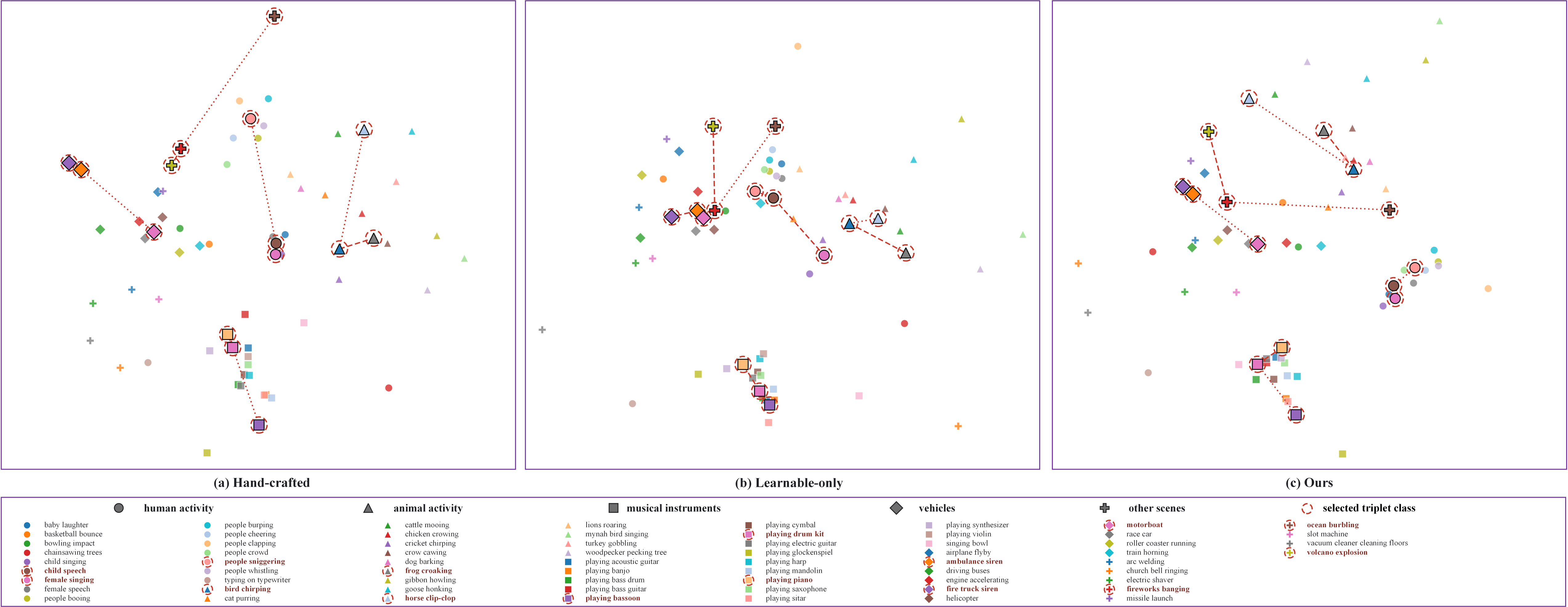}
    \caption{Semantic topology structure under different prompt strategies. 
    We derive t-SNE layouts from the inter-class affinity matrices
    of all 67 event categories from OV‑AVEBench~\cite{zhou2025towards}. 
    Marker shapes denote five semantic
    groups and colors distinguish individual categories. 
    Within each group, 
    red rings mark an anchor, 
    a semantically related category (long-dashed edge), 
    and a less related category (short-dashed edge). 
    }
    \label{fig:tsne}
    \vspace{-0.5em}
\end{figure*}

\begin{figure}[t]
    \centering
    \includegraphics[width=\linewidth]{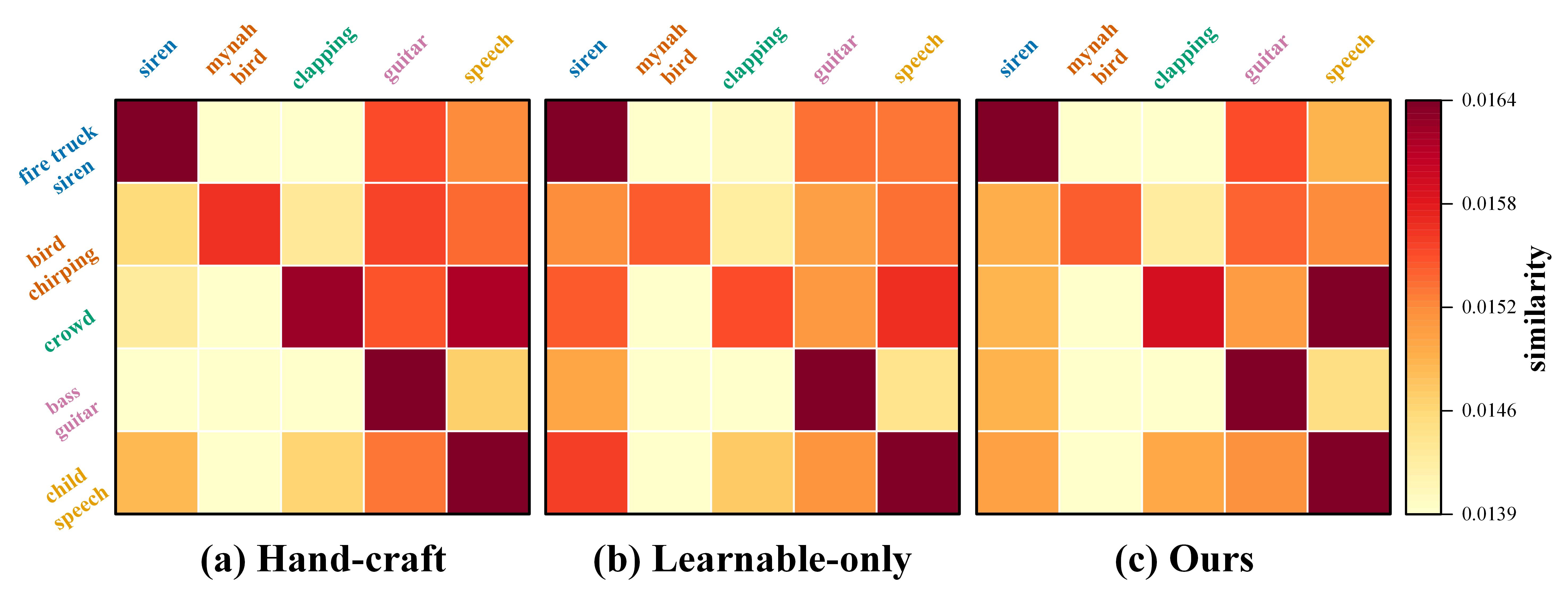}
    \caption{Pairwise semantic relation under different prompt strategies, where each row corresponds to a target event and each column to a candidate neighbor.}
    \label{fig:heatmap}
    \vspace{-1.5em}
\end{figure}

We provide cross‑validated evidence for prompt‑induced semantic topology drift from two complementary analytical perspectives.
Consistent observations from both Fig.~\ref{fig:tsne} and Fig.~\ref{fig:heatmap}
reveal that hand‑crafted prompts preserve intrinsic semantic structure of the category space. 
In contrast, unconstrained learnable prompts induce semantic topology drift and disrupt this structure. 
Our proposed STCR largely restores these required semantic relationships,
which is crucial for open‑vocabulary generalization.
More results and analyses are provided in \supplementary{Supplementary Materials}.

\subsubsection{Semantic Topology Structure Analysis}
As visualized in Fig.~\ref{fig:tsne},
hand‑crafted prompts yield a stable semantic structure and preserve reasonable local‑neighbor relationships across five semantic groups.
For instance, the category of \textit{ambulance siren} lies close to \textit{fire‑truck siren} 
while remaining distant from \textit{motorboat}.
Such consistent semantic patterns are also observed for the other four semantic groups.
In contrast,
learnable‑only prompts severely distort this topological organization.
For the five annotated triplets,
semantically related categories are repelled from their anchor, while weakly related ones are drawn nearer.
With the proposed regularization, 
semantic ordering is substantially improved, 
and the embedding topology moves closer to the hand‑crafted prompt.

\subsubsection{Pairwise Semantic Relation Analysis}

Fig.~\ref{fig:heatmap} 
examines pairwise semantic relations for five representative semantically‑associated category pairs within the affinity space.
Under hand‑crafted prompts, semantically‑matched pairs yield high similarity values, such as \emph{fire truck siren} and \emph{ambulance siren}, \emph{people crowd} and \emph{people clapping}, \emph{playing bass guitar} and \emph{playing electric guitar}, and \emph{child speech} and \emph{female speech}.
Learnable‑only prompting weakens their directed pairwise similarities, 
while several irrelevant off‑diagonal entries achieve higher similarity values.
Our STCR partially recovers these degraded pairwise similarities.

\subsection{Qualitative Results and Analysis}

\begin{figure*}[t]
\centering
\includegraphics[width=2.0\columnwidth]{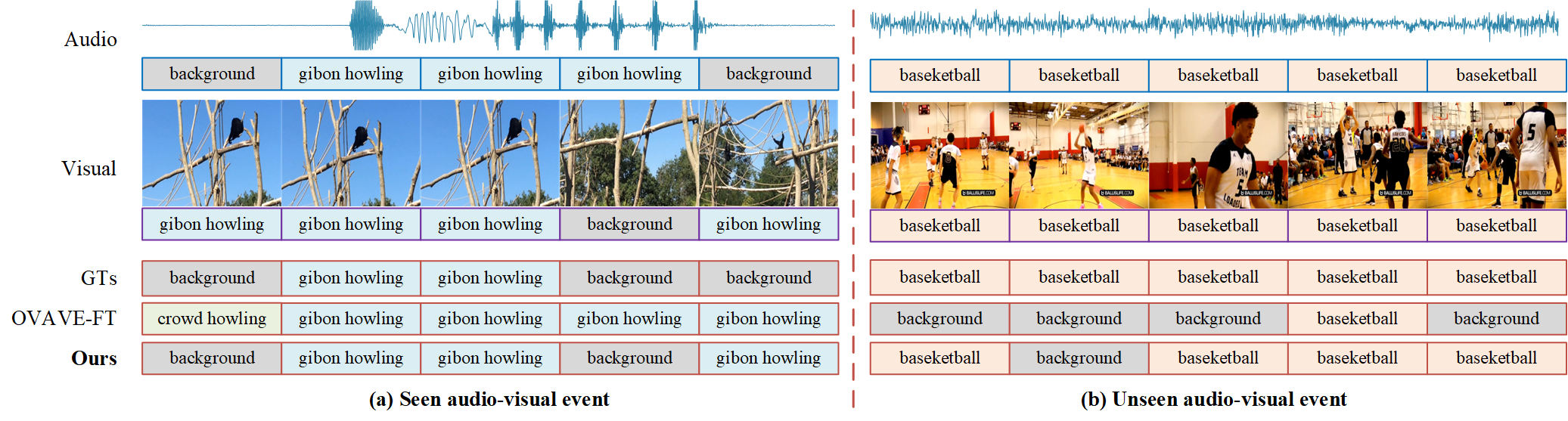}
\caption{Qualitative examples for seen and unseen audio-visual event categories.}
\label{fig6}
\end{figure*}

To qualitatively evaluate the temporal localization performance, 
we compare our method with the OVAVE-FT~\cite{zhou2025towards} on representative seen and unseen events. 
Fig.~\ref{fig6}(a) shows a representative example for the seen event  \emph{gibbon howling},
where audio and visual cues are temporally mismatched.
OVAVE-FT~\cite{zhou2025towards} misclassifies the initial background segment as the semantically similar category \emph{crowd howling} and incorrectly
propagates this event label to subsequent background segments. 
In contrast, 
our method correctly identifies background and yields accurate predictions for audio‑only segments with no corresponding visual cues.
Fig.~\ref{fig6}(b) shows localization results for the unseen event \emph{basketball}.
While OVAVE‑FT~\cite{zhou2025towards} correctly detects \emph{basketball} within only one segment and labels the rest as background,
our method successfully localizes this unseen event across most of its temporal span.
These qualitative results indicate that our method better distinguishes between semantically similar categories and achieves more precise event boundary localization for events.

\begin{table*}[t]
\centering
\caption{Ablation Studies on the core components of our framework. 
}
\begin{tabular}{ccc cccc cccc cccc}
\toprule
\multirow{2}{*}[-1ex]{{hand.}} & \multirow{2}{*}[-1ex]{{LCPT}} & \multirow{2}{*}[-1ex]{{STCR}} & 
\multicolumn{4}{c}{\textbf{Seen}} & 
\multicolumn{4}{c}{\textbf{Unseen}} & 
\multicolumn{4}{c}{\textbf{Total}} \\
\cmidrule(lr){4-7} \cmidrule(lr){8-11} \cmidrule(lr){12-15}
& & & \textit{Acc}. & \textit{Seg}. & \textit{Eve}. & \textit{Avg}. & \textit{Acc}. & \textit{Seg}. & \textit{Eve}. & \textit{Avg}. & \textit{Acc}. & \textit{Seg}. & \textit{Eve}. & \textit{Avg}. \\
\midrule
\cmark & \xmark& \xmark& {78.4} & {70.4} & {65.0} & {71.3} & {68.1} & {58.5} & {49.4} & {58.7} & {67.3} & {57.3} & {49.6} & {58.1} \\
\xmark & \cmark& \xmark& {78.9} & {70.2} & {64.8} & {71.3 }& {69.4} & {59.5} & {49.8} & {59.6} & {67.5} & {57.0} & {48.5} & {57.7} \\
\midrule
\rowcolor{gray!20}
\xmark& \cmark& \cmark&   {79.1} & {70.0} & {64.5} & {71.2} & {71.3} & {61.5} & {52.3} & {61.7} & {69.8} & {59.3} & {51.4} & {60.2}  \\
\midrule
\xmark& \cmark& \cmark $_\text{w/o novel}$&  {78.7} & {70.0} & {64.5} & {71.1} & {69.3} & {59.3} & {49.7} & {59.4} & {67.6} & {57.1} & {48.7} & {57.8}  \\
\xmark& \cmark& \cmark $_\text{w/o random}$&  {78.8} & {69.8} & {64.3} & {71.0} & {70.4} & {60.2} & {50.8} & {60.5} & {68.8} & {57.9} & {49.7} & {58.8}  \\
\bottomrule
\end{tabular}
\label{tab5}
\vspace{-1em}
\end{table*}

\subsection{Ablation Studies}

\subsubsection{Core components}
We conduct ablation experiments as summarized in Tab.~\ref{tab5} to assess the effectiveness of each core component within our framework, 
including LCPT and STCR with random category sampling and synthetic novel categories.

\paragraph{LCPT}
As reported in the first two rows,
the model with LCPT obtains average scores of 71.3\% and
59.6\% on the seen and unseen splits, respectively.
Compared with the hand-crafted prompts,
the seen-split performance stays unchanged,
whereas the unseen-split score rises by 0.9 \textit{pp}.
These observations indicate that LCPT adapts the text embeddings to
task-specific data and alleviates semantic mismatches between fixed templates and audio-visual event semantics.

\paragraph{STCR}
Comparing the $2$-nd and $3$-rd rows,
integrating STCR boosts average unseen-split and total-split
performance from 59.6\% and 57.7\% to 61.7\% and 60.2\%,
respectively, corresponding to gains of 2.1 and 2.5 \textit{pp}.
This validates that STCR relieves semantic topology drift caused by LCPT, enabling better generalization to unseen categories.

\paragraph{Novel Category Descriptions}
The row marked ``w/o novel'' presents results when synthetic novel categories are removed from STCR.
Removing these categories degrades the average score by 2.3 and
2.4 \textit{pp} on the unseen and total splits, respectively.
This outcome reveals that injecting novel category descriptions
simulates unseen events during training and improves recognition
of out-of-distribution event instances.

\paragraph{Random Embedding Sampling}
The row with ``w/o random'' replaces our random category sampling strategy with full pairwise computation across all categories.
Compared with the full model, removing random sampling reduces the average
score from 61.7\% to 60.5\% on the unseen split and from 60.2\% to
58.8\% on the total split, corresponding to drops of 1.2 and
1.4 \textit{pp}, respectively.
This result demonstrates the importance of stochastic category sampling for effective unseen-category generalization.

\begin{figure}[t]
\centering
\includegraphics[width=\columnwidth]{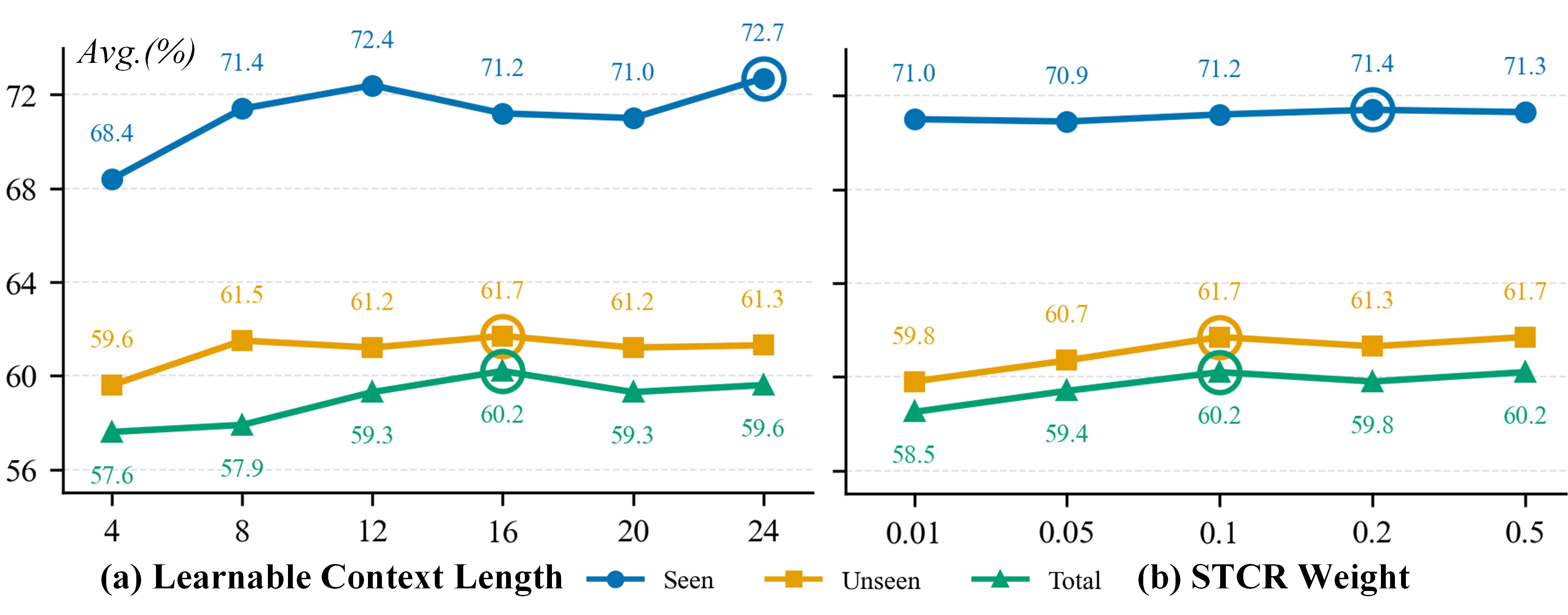}
\caption{
Ablation Studies about (a) the learnable context length $P$ in Eq.\ref{Eq3}
and (b) the regularization weight $\lambda$ in Eq.\ref{Eq9}.}
\label{fig:param_ablation}
\end{figure}

\subsubsection{Key hyper-parameters}
Fig.~\ref{fig:param_ablation} presents ablation results about the learnable context length $P$ in Eq.~\ref{Eq3}
and the regularization weight $\lambda$ in Eq.~\ref{Eq9}.
We conduct a grid search by varying one hyperparameter at a time while holding the other fixed.

\paragraph{Learnable Context Length}
Fig.~\ref{fig:param_ablation}(a) illustrates the performance across varying learnable context lengths
$P \in \{4,8,12,16,20,24\}$.
The total average score monotonically increases to its peak of 60.2\% at $P=16$ and declines thereafter.
The unseen-split performance also peaks at 61.7\% under this setting, while the seen-split score continuously improves with longer contexts.

\paragraph{Regularization Weight}
As shown in Fig.~\ref{fig:param_ablation}(b),
we further sweep $\lambda$ within $\{0.01,0.05,0.1,0.2,0.5\}$.
Insufficient regularization at  $\lambda=0.01$
yields a low Total average score of 58.5\%.
Performance peaks at 60.2\% for $\lambda=0.1$ and remains stable with larger weights, demonstrating robustness against regularization strength.
The unseen‑split average score also reaches its optimum of 61.7\% at $\lambda=0.1$, whereas the seen‑split performance peaks at 71.4\% when $\lambda=0.2$.
Accordingly, we select $\lambda=0.1$, the smallest weight achieving optimal performance on both total and unseen splits.

\section{Conclusion}

This paper presents a lightweight semantic structure-aligned prompt learning framework  tailored for open-vocabulary audio-visual event localization,
which preserves meaningful inter-category semantic relationships by aligning the underlying semantic topology. 
Our framework integrates  a learnable continuous prompt tuning module for automated prompt engineering and data-driven  cross‑modal alignment,
as well as a semantic topology consistency regularization module to alleviate the semantic projection shift problem.
Comprehensive experiments and in-depth analyses demonstrate the effectiveness of the proposed method and its strong generalization performance on unseen event categories.

\bibliographystyle{IEEEtran}
\bibliography{reference}

\end{document}